\documentclass[a4paper,11pt]{article}
\usepackage{jcappub} 
\usepackage[latin9]{inputenc}
\usepackage{textcomp}
\usepackage{amsmath}
\usepackage{amssymb}
\usepackage{color}

\makeatletter

\newcommand{\be}[1]{\begin{equation}\label{#1}}
\newcommand{\ee}{\end{equation}}
\newcommand{\ba}[1]{\begin{eqnarray}\label{#1}}
\newcommand{\ea}{\end{eqnarray}}

\makeatother

\begin{document} 
\title{Towards the Quantization of Eddington-inspired-Born-Infeld Theory}
\author[a,b,c,d]{Mariam Bouhmadi-L\'{o}pez}
\author[e,f]{and Che-Yu Chen}
\affiliation[a]{Departamento de F\'{i}sica, Universidade
da Beira Interior Rua Marqu\^{e}s D'\'Avila e Bolama, 6201-001 Covilh\~a, Portugal\\}
\affiliation[b]{Centro de Matem\'atica e Aplica\c{c}\~oes
da Universidade da Beira Interior (CMA-UBI), Rua Marqu\^{e}s D'\'Avila e Bolama, 6201-001 Covilh\~a, Portugal\\}
\affiliation[c]{Department of Theoretical Physics, University of the Basque Country
UPV/EHU, P.O. Box 644, 48080 Bilbao, Spain\\}
\affiliation[d]{IKERBASQUE, Basque Foundation for Science, 48011, Bilbao, Spain\\}
\affiliation[e]{Department of Physics, National Taiwan University, Taipei, Taiwan 10617\\}
\affiliation[f]{LeCosPA, National Taiwan University, Taipei, Taiwan 10617\\}
\emailAdd{mbl@ubi.pt (On leave of absence from UPV and IKERBASQUE)}
\emailAdd{b97202056@gmail.com}

\abstract{The quantum effects close to the classical big rip singularity within the Eddington-inspired-Born-Infeld theory (EiBI) are investigated through quantum geometrodynamics. It is the first time that this approach is applied to a modified theory constructed upon Palatini formalism. The Wheeler-DeWitt (WDW) equation is obtained and solved based on an alternative action proposed in Ref.~\cite{Delsate:2012ky}, under two different factor ordering choices. This action is dynamically equivalent to the original EiBI action while it is free of square root of the spacetime curvature. We consider a homogeneous, isotropic and spatially flat universe, which is assumed to be dominated by a phantom perfect fluid whose equation of state is a constant. We obtain exact solutions of the WDW equation based on some specific conditions. In more general cases, we propose a qualitative argument with the help of a Wentzel-Kramers-Brillouin (WKB) approximation to get further solutions. Besides, we also construct an \textit{effective} WDW equation by simply promoting the classical Friedmann equations. We find that for all the approaches considered, the DeWitt condition hinting singularity avoidance is satisfied. Therefore the big rip singularity is expected to be avoided through the quantum approach within the EiBI theory.
}

\maketitle
\flushbottom

\section{Introduction}
\label{sec:intro}
Undeniably, Einstein's theory of General Relativity (GR) has been an extremely successful theory for more than a century \cite{gravitation}. However, the theory is expected to break down at some points at very high energies where quantum effects are expected to become crucial, such as in the past expansion of the Universe where GR predicts a big bang singularity \cite{largescale}. On the other hand, several observations have found concrete evidences that the universe has entered a state of acceleration on its largest scale \cite{Perlmutter:1998np,Riess:1998cb}. Such an accelerating expansion can be fueled by an \textit{effective} dark energy whose equation of state is rather similar to that of a cosmological constant but allow the deviation from it to quintessence and phantom behaviors. The latter is quite interesting because it may give rise to future singularities \cite{Starobinsky:1999yw,Caldwell:2003vq,Caldwell:1999ew,Carroll:2003st,Chimento:2003qy,Dabrowski:2003jm,GonzalezDiaz:2003rf,GonzalezDiaz:2004vq,Nojiri:2005sx,BouhmadiLopez:2006fu,BouhmadiLopez:2004me}. Among them, the most destructive one is dubbed the big rip singularity \cite{Starobinsky:1999yw,Caldwell:2003vq,Caldwell:1999ew,Carroll:2003st,Chimento:2003qy,Dabrowski:2003jm,GonzalezDiaz:2003rf,GonzalezDiaz:2004vq,BouhmadiLopez:2004me}. It has been shown that if the Universe is dominated by a phantom dark energy with a constant equation of state, it will expand so violently that all bound structures will be ripped apart before a finite cosmic time when the singularity occurs. At the singularity, the size of the Universe and its expansion rate diverge, accompanied by the laceration of spacetime itself. Therefore, it is necessary to look for possible modified theories of gravity, which could explain the late time acceleration of the Universe. Furthermore, accompanied by some additional quantum effects, these modified theories of gravity may be able to smooth the singularities predicted in GR.

Recently, an alternative theory of gravity dubbed Eddington-inspired-Born-Infeld theory proposed in Ref.~\cite{Banados:2010ix}, pioneered in \cite{Deser:1998rj}, has attracted a lot of attention \cite{Bouhmadi-Lopez:2013lha,Scargill:2012kg,Avelino:2012ue,EscamillaRivera:2012vz,Yang:2013hsa,Du:2014jka,Wei:2014dka,Delsate:2012ky,Pani:2011mg,Pani:2012qb,Casanellas:2011kf,Avelino:2012ge,Avelino:2012qe,Harko:2013wka,Harko:2013aya,Sham:2013cya,Makarenko:2014lxa,Makarenko:2014nca,Odintsov:2014yaa,Pani:2012qd,Bouhmadi-Lopez:2014jfa,Makarenko:2014cca,Shaikh:2015oha,Jana:2015cha,Sotani:2015tya,Cho:2014xaa,Sotani:2014lua,Bouhmadi-Lopez:2014tna,Elizalde:2016vsd,Sotani:2015ewa,Chen:2015eha,Jimenez:2014fla,Cho:2013usa,Cho:2013pea,Cho:2012vg,Tamang:2015tmd,Potapov:2014iva,Izmailov:2015xsa,Bambi:2016xme,Olmo:2013gqa,Bambi:2015sla,Olmo:2015dba,Olmo:2015bya}. The EiBI theory has been shown to be able to cure the big bang singularity for a radiation dominated universe through a loitering effect and a bounce in the past, with the coupling constant $\kappa$ being positive and negative, respectively \cite{Banados:2010ix,Scargill:2012kg}. The ability of the theory to smooth other cosmological singularities in a phantom dominated universe has also been studied in our previous works \cite{Bouhmadi-Lopez:2013lha,Bouhmadi-Lopez:2014jfa,Bouhmadi-Lopez:2014tna,Chen:2015eha}. Unfortunately, we found that even though the EiBI theory can lead to the avoidance of the big bang and the alleviation of some smoother singularities, the big rip singularity is still unavoidable. 

As mentioned in the first paragraph, near the singular states, it is expected that some quantum effects should come into play. Even though so far there is not a fully consistent quantum theory of gravity, the framework of quantum cosmology in which a homogeneous, isotropic and spatially flat universe is considered can reduce the complexity of the quantization of GR. One can resort to the approach of quantum geometrodynamics in which the WDW equation plays a central role and where a canonical quantization of the fields is performed \cite{qgkiefer}. In this approach, the dynamics of the wave function of the Universe as a whole is determined by the WDW equation. If the solutions to the WDW equation satisfy the DeWitt (DW) condition \cite{DeWitt:1967yk}, which states that the wave function vanishes near the region corresponding to the classical singularity, it is expected that the singularity is avoided by quantum effects. In Refs.~\cite{Albarran:2016ewi,Albarran:2015cda,Albarran:2015tga,Bouhmadi-Lopez:2013tua,BouhmadiLopez:2009pu,Kamenshchik:2007zj,Dabrowski:2006dd,Kamenshchik:2013naa}, various dark energy related cosmological singularities which appear in the classical theory of GR have been shown to be cured or smoothed within the quantum framework. 

It is worth to be mentioned that although there have been some works investigating the quantum cosmology in the framework of modified theories of gravity \cite{Kamenshchik:2016rtr,Vakili:2009he,Vakili:2008uj,Shojai:2008er,capozziellogravity}, it is the first time that such an approach is applied to a theory constructed upon Palatini formalism. Because of the independence of the physical metric and the connection, the wave function contains two independent variables, one of which corresponds to the physical scale factor of the Universe and the other one corresponds to the physical connection components. It is the former that is responsible for the occurrence of the big rip singularity classically. Moreover, because of the complexity resulting from the square root structure of the EiBI action, we will consider an alternative action proposed in Ref.~\cite{Delsate:2012ky} to construct the Hamiltonian. We will also consider two different factor ordering choices to construct the WDW equations to see if our results depend on the factor ordering or not. The matter content is described by a phantom perfect fluid whose equation of state $w$ is a constant, i.e., $w<-1$. Some exact solutions can be obtained based on the assumption that the physical scale factor and the connections are related through a classical motion equation. For more general circumstances in which these two quantities are treated independently, a qualitative argument is proposed with the help of the WKB approximations and we show that the wave functions vanish near the region corresponding to the classical big rip singularity. Furthermore, we also construct another WDW equation by promoting the Friedmann equations of the theory. According to the DW condition for singularity avoidance, we find that for all the approaches considered in this work, the big rip singularity is expected to be avoided within a quantum realm.

This paper is outlined as follows. In section~\ref{sec:eibiclassical}, we briefly review the classical EiBI phantom model in which the phantom energy has a constant equation of state. This phantom energy leads to a big rip singularity in the EiBI theory. In section~\ref{seclangran}, we construct the WDW equation by considering an alternative action proposed in Ref.~\cite{Delsate:2012ky} under two different factor ordering choices. We find that the big rip singularity is hinted to be avoided according to the DW condition. In section~\ref{sec:second}, we construct an effective WDW equation by simply promoting the Friedmann equations of the EiBI theory. The same conclusion, i.e., singularity avoidance, can be reached. We finally present our conclusions in section~\ref{sec:conclusion}. Some cumbersome but necessary calculations are presented in the appendices.

\section{The EiBI phantom model: constant equation of state}
\label{sec:eibiclassical}
We start reviewing the EiBI scenario whose gravitational action is \cite{Banados:2010ix}
\begin{equation}
S_{EiBI}=\frac{2}{\kappa}\int d^4x\Big[\sqrt{|g_{\mu\nu}+\kappa R_{\mu\nu}(\Gamma)|}-\lambda\sqrt{-g}\Big]+S_M(g).
\label{actioneibi}
\end{equation}
On the previous expression $|g_{\mu\nu}+\kappa R_{\mu\nu}(\Gamma)|$ is the determinant of the rank two tensor $g_{\mu\nu}+\kappa R_{\mu\nu}(\Gamma)$. The theory is formulated within a Palatini formalism, in which the metric $g_{\mu\nu}$ and the connection $\Gamma$ are treated as independent variables. In addition, $R_{\mu\nu}(\Gamma)$ is chosen to be the symmetric part of the Ricci tensor and the connection is assumed to be torsionless. Note that $g$ is the determinant of the metric and $S_M$ stands for the matter Lagrangian, where matter is assumed to be coupled covariantly to the metric $g$ only. In addition, $\lambda$ is a dimensionless constant which relates to an effective cosmological constant of the theory at low curvature through $\Lambda\equiv(\lambda-1)/\kappa$. The parameter $\kappa$ is a constant characterizing the theory. In this paper, we will work with Planck units $8\pi G=1$ and set the speed of light to $c=1$.

Variation of action \eqref{actioneibi} with respect to the connection and the metric $g_{\mu\nu}$ leads to the field equations \cite{Banados:2010ix}
\begin{subequations}\label{eq12}
\begin{align}
\lambda q_{\mu\nu}&=g_{\mu\nu}+\kappa R_{\mu\nu},\label{eq1}\\
q^{\mu\nu}&=\tau(g^{\mu\nu}-\frac{\kappa}{\lambda}T^{\mu\nu}),\label{eq2}
\end{align}
\end{subequations}
respectively, where $q_{\mu\nu}$ is the auxiliary metric compatible with the connection. Furthermore, $q^{\mu\nu}$ and $q$ represent the inverse matrix and the determinant of $q_{\mu\nu}$, respectively. The energy momentum tensor is defined by
\begin{equation}
T^{\mu\nu}=\frac{1}{\sqrt{-g}}\frac{\delta S_M}{\delta g_{\mu\nu}},
\label{energymomentumT}
\end{equation}
and $\tau\equiv\sqrt{g/q}$. It can be shown that the tensor $q_{\mu\nu}$ is the metric compatible with the connection $\Gamma$, i.e., $\nabla_{\alpha}q_{\mu\nu}=0$ where $\nabla$ is a covariant derivative constructed from the connection $\Gamma$. In addition, the dependence of $q_{\mu\nu}$ on the physical metric $g_{\mu\nu}$ and the dependence of $q_{\mu\nu}$ on $T_{\mu\nu}$ are determined by the field equations \eqref{eq12}. We review the derivation of the field equations \eqref{eq12} and how the compatibility of $q_{\mu\nu}$ and $\Gamma$ arises on the appendix \ref{app0}. On the particular case of a homogeneous and isotropic universe, the most general equations of motion for an arbitrary perfect fluid were obtained in Ref.~\cite{Bouhmadi-Lopez:2014jfa}. In particular on that paper, we obtained the Friedmann equation corresponding to both metrics. As both of them depends on the energy density, there is as well a relation between the physical Hubble rate and the auxiliary Hubble rate. However, this relation is not that enlightening and quite complicated, that is the reason why we omitted it on that paper and on the current one as well. Here we follow the same notation as in Ref.~\cite{Delsate:2012ky}, that is,  $q_{\mu\nu}$ is defined by the auxiliary metric used in Ref.~\cite{Banados:2010ix} divided by a factor of $\lambda$. In addition, we will restrict our analysis to positive $\kappa$, in order to avoid the imaginary effective sound speed instabilities usually present in the EiBI theory with negative $\kappa$ \cite{Avelino:2012ge}.

For a homogeneous, isotropic and spatially flat universe, we will consider the Friedmann-Lema\^itre-Robertson-Walker metric (FLRW)
\begin{equation}
g_{\mu\nu}dx^{\mu}dx^{\nu}=-dt^2+a(t)^2d\vec{x}^2,
\end{equation}
where $t$ is the cosmic time and $a(t)$ is the physical scale factor. In Ref.~\cite{Bouhmadi-Lopez:2013lha,Bouhmadi-Lopez:2014jfa}, we have shown that if the Universe is filled with a phantom energy, which is described by a perfect fluid with a constant equation of state $w<-1$, the Hubble rate $H\equiv\dot{a}/a$ and its cosmic derivative have the following approximated behaviors at late time
\begin{eqnarray}
H^2&\approx&\frac{4\sqrt{|w|^3}}{3(3w+1)^2}\rho\rightarrow\infty,\nonumber\\
\dot{H}&\approx&\frac{2\sqrt{|w|^3}}{(3w+1)^2}|1+w|\rho\rightarrow\infty.
\end{eqnarray}
Note that the dot denotes the derivative with respect to the cosmic time $t$. The phantom energy density $\rho$ and pressure $p=w\rho$ diverge as well. At the same time, the scale factor $a(t)$ blows up. This singular state happens at a finite cosmic time $t$ and corresponds to the big rip singularity in the EiBI theory \cite{Bouhmadi-Lopez:2013lha,Bouhmadi-Lopez:2014jfa}. Note that in Ref.~\cite{Bouhmadi-Lopez:2014jfa}, we also showed that there is no singularity of the auxiliary metric $q_{\mu\nu}$ when an appropriate rescaled cosmic time and an auxiliary scale factor are chosen such that the auxiliary metric takes the form of a FLRW metric. In fact, when the physical metric $g_{\mu\nu}$ faces  the big rip, the auxiliary metric compatible with the physical connection is asymptotically de Sitter and therefore well defined \cite{Bouhmadi-Lopez:2014jfa}.

\section{First approach to quantize: Using the effective Lagrangian approach}\label{seclangran}
\subsection{The effective Lagrangian}
As the big rip singularity is unavoidable in the EiBI phantom model, it is natural to ask whether the quantum effect can shed some light over preventing this cosmic doomsday. We will consider a quantum geometrodynamical approach in which the WDW equation plays a central role. However, the complexity resulting from the square root of the curvature in the action \eqref{actioneibi} is not easy to overcome. Therefore, in this work we will resort to an alternative action, which is dynamically equivalent to the original EiBI action.

In Ref.~\cite{Delsate:2012ky}, the authors showed that the field equations \eqref{eq12} imply that Einstein tensor for $q_{\mu\nu}$ satisfies:
\begin{eqnarray}
{G^{\mu}}_{\nu}[q]&\equiv&q^{\mu\alpha}R_{\alpha\nu}-\frac{1}{2}{\delta^\mu}_\nu q^{\alpha\beta}R_{\beta\alpha}\nonumber\\
&=&\frac{1}{\lambda}\Big[\tau {T^{\mu}}_{\nu}+{\delta^{\mu}}_{\nu}\Big(\frac{\lambda}{\kappa}(\tau-1)-\frac{1}{2}\tau T\Big)\Big]-\Lambda{\delta^{\mu}}_{\nu},
\label{einsteintensorq}
\end{eqnarray}
where ${T^{\mu}}_{\nu}=T^{\mu\alpha}g_{\alpha\nu}$ and $T={T^{\mu}}_{\mu}$. In Ref.~\cite{Delsate:2012ky}, the authors also showed that the field equations \eqref{eq12} can be obtained from the following alternative action
\begin{equation}
S_{a}=\lambda\int d^4x\sqrt{-q}\Big[R(\Gamma)-\frac{2\lambda}{\kappa}+\frac{1}{\kappa}(q^{\alpha\beta}g_{\alpha\beta}-2\tau)\Big]+S_M(g),
\label{actionalternative}
\end{equation}
where $R(\Gamma)\equiv q^{\alpha\beta}R_{\beta\alpha}(\Gamma)$. This action is similar to a bi-gravity action without dynamics for $g_{\mu\nu}$ \cite{Delsate:2012ky}. More precisely, the variation of action \eqref{actionalternative} in terms of $g$ gives equation \eqref{eq2}, and the variation in terms of $q$ gives the Einstein equation of $q$, that is, Eq.~\eqref{einsteintensorq}, which implies the field equation \eqref{eq1}. Therefore, the actions \eqref{actioneibi} and \eqref{actionalternative} are equivalent dynamically because they give the same classical field equations \cite{Delsate:2012ky}. However, the action \eqref{actionalternative} has a similar form with the Einstein Hilbert action in the sense that it is linear on $R(\Gamma)$, and most importantly it does not contain a square root of the curvature; i.e., a square root involving second derivatives of the scale factor of the metric compatible with $\Gamma$. Therefore, it is easier to get the classical Hamiltonian, which is very important to get the WDW equation, for this model.

The starting point is the homogeneous and isotropic ansatz of the Universe, therefore
\begin{eqnarray}
g_{\mu\nu}dx^{\mu}dx^{\nu}&=&-N(t)^2 dt^2+a(t)^2d\vec{x}^2,\nonumber\\
q_{\mu\nu}dx^{\mu}dx^{\nu}&=&-M(t)^2 dt^2+b(t)^2d\vec{x}^2.\nonumber
\end{eqnarray}
$N(t)$ and $M(t)$ are the lapse functions of $g_{\mu\nu}$ and $q_{\mu\nu}$, respectively. These lapse functions are included from now on for the sake of completeness. Similarly, $a$ and $b$ correspond to the scale factor of each metric. The Ricci scalar constructed solely from $q$ is 
\begin{equation}
R(\Gamma)\equiv q^{\alpha\beta}R_{\beta\alpha}(\Gamma)=\frac{6}{M^2}\Big[\frac{\ddot{b}}{b}+\Big(\frac{\dot{b}}{b}\Big)^2-\frac{\dot{b}}{b}\frac{\dot{M}}{M}\Big].
\end{equation}
Then, we consider the simplest case in which the matter component is described by a perfect fluid with a given equation of state. Therefore, the matter can be purely described by a single variable $a$, the scale factor of $g_{\mu\nu}$.

After integrations by part, the FLRW Lagrangian constructed from the action \eqref{actionalternative}, $S_a=v_0\int dt\mathcal{L}$, can be rewritten as
\begin{equation}
\mathcal{L}=\lambda Mb^3\Big[-\frac{6{\dot{b}}^2}{M^2b^2}-\frac{2\lambda}{\kappa}+\frac{1}{\kappa}\Big(\frac{N^2}{M^2}+3\frac{a^2}{b^2}-2\frac{Na^3}{Mb^3}\Big)\Big]-2\rho(a)Na^3,
\label{LA}
\end{equation}
where $v_0$ corresponds to the spatial volume after a proper compactification for spatially flat sections \cite{qgkiefer}.

\subsection{The Euler-Lagrange equation of the system}
In this subsection, we calculate the Euler-Lagrange equation from the Lagrangian \eqref{LA}. The Euler-Lagrange equation of $N$ and $a$ are
\begin{subequations}\label{eqnma}
\begin{align}
\lambda\frac{Nb^3}{Ma^3}&=\lambda+\kappa\rho,\label{N}\\
3\lambda bM&=3aN(\lambda+\kappa\rho)+\kappa a^2N\frac{d\rho}{da},\label{awithout}
\end{align}
respectively. The relation between $d\rho/da$ and pressure $p$ is given by the conservation equation $d\rho/da=-3(\rho+p)/a$. Inserting the conservation equation into Eq.~\eqref{awithout}, it becomes
\begin{equation}
\lambda\frac{bM}{aN}=\lambda-\kappa p.
\label{a}
\end{equation}
\end{subequations}
Actually, equations \eqref{N} and \eqref{a} can be rewritten as
\begin{subequations}\label{na}
\begin{align}
\frac{(\lambda-\kappa p)^3}{\lambda+\kappa\rho}&=\lambda^2\frac{M^4}{N^4}=U^2,\label{UUU}\\
(\lambda+\kappa\rho)(\lambda-\kappa p)&=\lambda^2\frac{b^4}{a^4}=V^2\label{VVV},
\end{align}
\end{subequations}
where $U$ and $V$ are the notations representing $q_{00}$ and $q_{ij}$ in Ref.~\cite{Banados:2010ix}. Near the big rip singularity, the energy density is dominated by phantom energy whose scale factor dependence can be described as $a^{\epsilon}$, with $\epsilon$ positive and given by $\epsilon\equiv-3(1+w)$. Therefore, both $\rho$ and $p$ behave like $a^\epsilon$ near the big rip singularity. Thus, from Eq.~\eqref{VVV}, we have $b^4\propto a^{4+2\epsilon}$ asymptotically. 

The Euler-Lagrange equation of $M$ leads to the Friedmann equation of $b$:
\begin{equation}
\kappa\Big(\frac{\dot{b}}{b}\Big)^2=M^2\Big(\frac{1}{6}\frac{N^2}{M^2}-\frac{1}{2}\frac{a^2}{b^2}+\frac{\lambda}{3}\Big),
\label{feqb}
\end{equation}
and the Euler-Lagrange equation of $b$ is the Raychaudhuri equation of $b$:
\begin{equation}
\kappa\frac{\ddot{b}}{b}=N^2\Big(\frac{\lambda}{3}\frac{M^2}{N^2}-\frac{1}{3}\Big)+\kappa\frac{\dot{b}}{b}\frac{\dot{M}}{M}.
\label{reqb}
\end{equation} 
If we use Eqs.~\eqref{na} to replace $N^2/M^2$ and $a^2/b^2$ in Eq.~\eqref{feqb}, Eq.~\eqref{feqb} can be rewritten as
\begin{equation}
\kappa\Big(\frac{\dot{b}}{b}\Big)^2=\lambda M^2\Big[\frac{1}{3}+\frac{\kappa\rho+3\kappa p-2\lambda}{6\sqrt{(\lambda+\kappa\rho)(\lambda-\kappa p)^3}}\Big].
\end{equation}
One can see that the auxiliary Hubble rate defined by $H_q$ in our previous work (Eq.~$(2.10)$ in Ref.~\cite{Bouhmadi-Lopez:2014jfa}) is recovered when we choose the following lapse function for the auxiliary coordinate system:
\begin{equation*}
M^2=1/\lambda.
\end{equation*}
Therefore, it can be easily seen that if $w<-1$ and is a constant, $\dot{b}/b$ approaches a constant when $\rho$ goes to infinity. This corresponds to a de Sitter stage of the auxiliary metric as mentioned in the end of section.~\ref{sec:eibiclassical} and fully proven in Ref.~\cite{Bouhmadi-Lopez:2014jfa}.

 \subsection{The Hamiltonian and the WDW equations}
From the Lagrangian \eqref{LA}, one can see that the conjugate momenta of $a$, $N$ and $M$ are zero. However, the conjugate momenta of $b$ is not and reads
\begin{equation}
p_b=\frac{\partial\mathcal{L}}{\partial\dot{b}}=-\frac{12\lambda b}{M}\dot{b},
\label{pb}
\end{equation}
and the Hamiltonian is
\begin{eqnarray}
\mathcal{H}_1&=&\dot{b}p_b-\mathcal{L}\nonumber\\
&=&-\frac{M}{24\lambda b}p_b^2+\frac{2\lambda^2}{\kappa}Mb^3-\frac{\lambda}{\kappa}b^3\frac{N^2}{M}-\frac{3\lambda}{\kappa}Mba^2+\frac{2Na^3}{\kappa}(\lambda+\kappa\rho).
\label{Hamiltonian}
\end{eqnarray}
It can be shown that by using the constraint equations \eqref{N} and \eqref{feqb} we obtain $\mathcal{H}_1=0$. Therefore, we can construct another Hamiltonian $\mathcal{H}_2$ by using again the constraint Eq.~\eqref{N} to substitute $N/M$ in $\mathcal{H}_1$ by $\kappa\rho$, $a$, and $b$:
\begin{equation}
\mathcal{H}_2=M\Big[-\frac{p_b^2}{24\lambda b}+\frac{2\lambda^2}{\kappa}b^3+\frac{1}{\kappa\lambda}(\lambda+\kappa\rho(a))^2\frac{a^6}{b^3}-\frac{3\lambda}{\kappa}ba^2\Big].
\label{h3}
\end{equation}
Just like the case in $\mathcal{H}_1$, the second Hamiltonian $\mathcal{H}_2$ fulfills $\mathcal{H}_2=0$ if the constraints equations \eqref{N} and \eqref{feqb} are assumed simultaneously. We would like to clarify the following: there is an alternative way of showing that both Hamiltonians vanish. The Hamiltonians $\mathcal{H}_1$ and $\mathcal{H}_2$ are independent of time, so $\partial\mathcal{H}_i/\partial t =0$ where $i=1,2$. Therefore, these Hamiltonians are conserved. In addition, as they are zero on shell and continuous, they will vanish everywhere.


Because we only use the constraint Eq.~\eqref{N} to write down the Hamiltonian $\mathcal{H}_2$ from $\mathcal{H}_1$, without using Eq.~\eqref{a} which relates $a$ and $b$ in the classical regime, the variables $a$ and $b$ in $\mathcal{H}_2$ are treated as independent variables only at the quantum level. Notice that we do not use Raychaudhuri equation in the classical Hamiltonian in GR either.

As there is no singularity at $b=0$ for the model, we can safely rescale the Hamiltonian as
\begin{eqnarray}
b^3\mathcal{H}_2&=&M\Big[-\frac{b^2p_b^2}{24\lambda}+\frac{2\lambda^2}{\kappa}b^6+\frac{1}{\kappa\lambda}(\lambda+\kappa\rho(a))^2a^6-\frac{3\lambda}{\kappa}a^2b^4\Big]\nonumber\\
&=&0.
\end{eqnarray}
Then, we can write down the WDW equation by choosing the following factor ordering:
\begin{equation}
b^2p_b^2=-\hbar^2\Big(b\frac{\partial}{\partial b}\Big)\Big(b\frac{\partial}{\partial b}\Big)=-\hbar^2\Big(\frac{\partial}{\partial x}\Big)\Big(\frac{\partial}{\partial x}\Big),
\end{equation}
where $x=\ln(\sqrt{\lambda}b)$. Therefore, the WDW equation reads:
\begin{equation}
\Big[\frac{\partial^2}{\partial x^2}+V_1(a,x)\Big]\Psi(a,x)=0,
\label{wdw1}
\end{equation}
where 
\begin{equation}
V_1(a,x)=\frac{24}{\kappa\hbar^2}\Big[2\textrm{e}^{6x}-3a^2\textrm{e}^{4x}+(\lambda+\kappa\rho(a))^2a^6\Big].
\label{VV}
\end{equation}
Next, we rewrite the potential $V_1(a,x)$ as
\begin{equation}
V_1(a,x)=\frac{24}{\kappa\hbar^2}\textrm{e}^{6x}[2-3\delta+(\lambda+\kappa\rho(a))^2\delta^3],
\label{26}
\end{equation}
where $\delta\equiv a^2\textrm{e}^{-2x}$. Near the classical big rip singularity where $a\rightarrow\infty$, the behavior of the potential can be classified as follows:
\begin{itemize}
\item If $a^2$ diverges slower than $\textrm{e}^{2x}$, i.e., $\delta\rightarrow 0$, the second term in the bracket in \eqref{26} is negligible compared with the first term. However, whether the first term dominates over the third term depends on the exact form of $\rho(a)$ and $\delta$. In either cases, the potential reaches positive infinite values when both $a$ and $x$ go to infinity.
\item If $a^2$ diverges faster than $\textrm{e}^{2x}$, i.e., $\delta\rightarrow\infty$, the potential can be approximated as
\begin{equation}
V_1(a,x)\approx\frac{24}{\kappa\hbar^2}(\lambda+\kappa\rho(a))^2a^6,
\label{27}
\end{equation}
when $a$ goes to infinity.
\item If $a^2$ diverges comparably with $\textrm{e}^{2x}$, the potential can also be approximated as in Eq.~\eqref{27} because the phantom energy density blows up when $a\rightarrow\infty$.
\end{itemize}
Therefore, we find that the potential $V_1(a,x)$ goes to positive infinity when $a\rightarrow\infty$ for all values of $x$.

As a guiding example, we temporarily assume that two scale factors $a$ and $b$ are related through Eq.~\eqref{VVV}. Note that this relation is a consequence of the assumption of the classical equations of motion \eqref{N} and \eqref{a}. If we further assume that the energy density and the pressure of the phantom energy behave like $a^{\epsilon}$ near the singularity, the scale factor $b$ and $a$ are related through $b^4\propto a^{4+2\epsilon}$ asymptotically. One can find that the potential $V_1(a,x)$ in \eqref{VV} is dominated by the first term. This belongs to the first case in the above qualitative discussions. Therefore, the WDW equation \eqref{wdw1} can be approximated as
\begin{equation}
\Big(\frac{d^2}{dx^2}+\frac{48}{\kappa\hbar^2}\textrm{e}^{6x}\Big)\Psi(x)=0,
\label{diff1}
\end{equation}
when $a$ and $x$ approach infinity. The solution is \cite{mathhandbook}
\begin{equation}
\Psi(x)=C_1J_0(A_1\textrm{e}^{3x})+C_2Y_0(A_1\textrm{e}^{3x}),
\end{equation}
and consequently when $x\rightarrow\infty$, its asymptotic behavior reads \cite{mathhandbook}
\begin{equation}
\Psi(x)\approx\sqrt{\frac{2}{\pi A_1}}\textrm{e}^{-3x/2}\Big[C_1\cos{\Big(A_1\textrm{e}^{3x}-\frac{\pi}{4}\Big)}+C_2\sin{\Big(A_1\textrm{e}^{3x}-\frac{\pi}{4}\Big)}\Big],
\end{equation}
where 
\begin{equation}
A_1\equiv\frac{4}{\sqrt{3\kappa\hbar^2}}.
\end{equation}
Here $J_\nu(x)$ and $Y_\nu(x)$ are Bessel function of first kind and second kind, respectively \cite{mathhandbook}. Therefore, the wave function $\Psi(x)$ approaches zero when $a$ as well as $x$ go to infinity. According to the DeWitt criterium for singularity avoidance, the big rip singularity is expected to be avoided.

However, in general the variables $a$ and $b$ (or $x$) should be treated independently in the quantum realm. The big rip singularity happens when $a$ goes to infinity, instead of $b$. Therefore, on the next subsection, we will go back to the WDW equation \eqref{wdw1} and verify that the wave function vanishes when $a$ approaches infinity for all values of $x$, with the help of a WKB approximation.

\subsection{The WKB approximation}\label{wkbsec}
In this subsection, we turn to solve the complete partial differential equation \eqref{wdw1}. Actually, the variable $a$ is not dynamical so we can regard this equation as an ordinary differential equation of $x$, leaving $a$ as a constant. To solve the differential equation, we apply a WKB approximation \cite{Albarran:2015cda,Albarran:2015tga,mathhandbook2}:

For the following differential equation \cite{Albarran:2015cda,Albarran:2015tga}
\begin{equation}
\{\partial_x^2+\mathcal{M} g(x)\}\Psi(x)=0,
\label{eqtest}
\end{equation}
where $\mathcal{M}$ is a constant, the first order WKB approximated solution reads \cite{Albarran:2015tga,mathhandbook2}:
\begin{equation}
\Psi(x)\approx g(x)^{-\frac{1}{4}}\Big[C_1e^{h(x)}+C_2e^{-h(x)}\Big],
\end{equation}
where 
\begin{equation}
h(x)=\int \sqrt{-\mathcal{M} g(x)}dx.
\end{equation}
Therefore, the first order WKB approximated solution to Eq.~\eqref{wdw1} is 
\begin{equation}
\Psi(a,x)\approx S(a,x)^{-\frac{1}{4}}\textrm{exp}\Big\{\pm\frac{2}{\hbar}\sqrt{\frac{6}{\kappa}}i\int^x\sqrt{S(a,x')}dx'\Big\},
\label{solution}
\end{equation}
where 
\begin{equation}
S(a,x)=2\textrm{e}^{6x}-3a^2\textrm{e}^{4x}+K^2(a)a^6,
\end{equation}
and
\begin{equation}
K(a)=\lambda+\kappa\rho(a).
\end{equation}
In summary, because the potential $V_1(a,x)$ goes to positive infinity when $a\rightarrow\infty$ for all values of $x$ as shown qualitatively in the previous subsection, the integral 
\begin{equation}
I\equiv\int^x\sqrt{S(a,x')}dx',\label{integrals}
\end{equation}
whose exact solution is given on the appendix\footnote{We have left the cumbersome but important calculations to two appendices to make the reading of the paper easier.}\ref{app1}, is real when $a$ gets large. The pre-factor $S(a,x)^{-1/4}$ is a decaying function and vanishes in the same limit. Hence, we can claim that the wave function $\Psi(a,x)$ vanishes for all values of $x$ when $a\rightarrow\infty$. According to the DeWitt criterium for singularity avoidance, the big rip singularity is expected to be avoided within this approach.

\subsubsection{Validity of the WKB approximation}
For the differential equation of the form in Eq.~\eqref{eqtest}, the validity of the first order WKB approximation is ensured by the following inequality (c.f. for example the appendix in \cite{Albarran:2015tga}):
\begin{equation}
q(x)\equiv\frac{1}{\mathcal{M}}\Bigg|\frac{5[g'(x)]^2-4[g''(x)][g(x)]}{16[g(x)]^3}\Bigg|\ll 1.
\end{equation}
In our WDW equation \eqref{wdw1}, this condition corresponds to
\begin{equation}
\Bigg|\frac{3[3-4\delta+3\delta^2-6\delta^3(\lambda+\kappa\rho(a))^2+4\delta^4(\lambda+\kappa\rho(a))^2]}{\textrm{e}^{6x}(2-3\delta+\delta^3(\lambda+\kappa\rho(a))^2)^3}\Bigg|\ll\frac{24}{\kappa\hbar^2}.
\end{equation}
It can be easily seen that this condition is satisfied when $a\rightarrow\infty$ for any value of $\delta$. This justifies the validity of the WKB approximations.

\subsection{Second factor ordering procedure}
From the Hamiltonian \eqref{h3} we can choose another factor ordering
\begin{equation}
\frac{p_b^2}{b}=-\hbar^2\Big(\frac{1}{\sqrt{b}}\frac{\partial}{\partial b}\Big)\Big(\frac{1}{\sqrt{b}}\frac{\partial}{\partial b}\Big),
\label{pbb}
\end{equation}
and write the corresponding WDW equation by introducing a new variable $y\equiv(\sqrt{\lambda}b)^{3/2}$ as follows
\begin{equation}
\Big[\frac{\partial^2}{\partial y^2}+V_2(a,y)\Big]\Psi(a,y)=0,
\label{eqan}
\end{equation}
where 
\begin{equation}
V_2(a,y)=\frac{32}{3\kappa\hbar^2}y^2\Big[2-3\eta+(\lambda+\kappa\rho(a))^2\eta^3\Big],
\label{V1YY}
\end{equation}
and $\eta\equiv a^2y^{-4/3}$. Before proceeding further, we highlight that this quantization is based on the Laplace-Beltrami operator which is the Laplacian operator in minisuperspace \cite{qgkiefer}. This operator depends on the number of degrees of freedom involved. For the case of a single degree of freedom, it can be written as in Eq.~\eqref{pbb} (c.f. for example \cite{Albarran:2015tga}). 

The behavior of $V_2(a,y)$ given in \eqref{V1YY} near the classical big rip singularity where $a\rightarrow\infty$ can be classified as follows:
\begin{itemize}
\item If $a^2$ diverges slower than $y^{4/3}$, i.e., $\eta\rightarrow 0$, the second term of $V_2(a,y)$ is negligible compared with the first term. However, whether the first term dominates over the third term depends on the exact form of $\rho(a)$ and $\eta$. In either cases, the potential reaches positive infinite values when both $a$ and $y$ go to infinity.
\item If $a^2$ diverges faster than $y^{4/3}$, i.e., $\eta\rightarrow\infty$, the potential can be approximated as
\begin{equation}
V_2(a,y)\approx\frac{32}{3\kappa\hbar^2}(\lambda+\kappa\rho(a))^2\eta^{\frac{3}{2}}a^3,
\label{40}
\end{equation}
when $a$ goes to infinity.
\item If $a^2$ diverges comparably with $y^{4/3}$, the potential can also be approximated as in Eq.~\eqref{40} because the phantom energy density blows up when $a\rightarrow\infty$.
\end{itemize}
Therefore, we find that the potential $V_2(a,y)$ goes to positive infinity when $a\rightarrow\infty$ for all values of $y$. Qualitatively, following the same arguments as those presented in subsection \ref{wkbsec}, we can claim that the wave function $\Psi(a,y)$ vanishes when $a\rightarrow\infty$. Hence, the big rip singularity is expected to be avoided as well for this factor ordering.

Alternatively, we can temporarily assume that the two scale factors $a$ and $b$ are related through Eq.~\eqref{VVV}. Therefore, by reminding that the energy density and pressure of the phantom fluid behave like $a^{\epsilon}$ near the singularity, we reach the conclusion that the scale factor $b$ and $a$ are related through $b^4\propto a^{4+2\epsilon}$ asymptotically. The potential $V_2(a,y)$ in Eq.~\eqref{V1YY} is dominated then by its first term. This belongs to the first case in the above qualitative discussions. Therefore, the WDW equation \eqref{eqan} can be approximated as
\begin{equation}
\Big(\frac{d^2}{dy^2}+\frac{64}{3\kappa\hbar^2}y^2\Big)\Psi(y)=0,
\label{diff2}
\end{equation}
when $a$ and $y$ approach infinity. The solution of the previous equation reads\cite{mathhandbook}
\begin{equation}
\Psi(y)=C_1\sqrt{y}J_{1/4}(A_1y^2)+C_2\sqrt{y}Y_{1/4}(A_1y^2),
\end{equation}
and when $y\rightarrow\infty$,
\begin{equation}
\Psi(y)\approx\sqrt{\frac{2}{\pi A_1y}}\Big[C_1\cos{\Big(A_1y^2-\frac{3\pi}{8}\Big)}+C_2\sin{\Big(A_1y^2-\frac{3\pi}{8}\Big)}\Big].
\end{equation}
Therefore, the wave function $\Psi(y)$ approaches zero when $a$ as well as $y$ go to infinity. According to the DeWitt criterium for singularity avoidance, the big rip singularity is expected to be avoided in this case.

The lesson we have learnt from these two factor orderings quantization is that our results seem to be independent of the chosen factor orderings.

\section{An alternative ``phenomenological'' quantization approach}\label{sec:second}
In this section, we introduce an alternative quantization approach, which is entirely independent of the quantization method mentioned previously, to investigate the quantum version of the EiBI setup. This approach is simply intended to construct an effective WDW equation by promoting the Friedmann equations in the classical theory rather than the Lagrangian. The definition of the conjugate momenta is inspired in GR. Although this approach seems rather phenomenological and not well-motivated as compared with the previous quantization approach, it can provide an additional confirmation on the robustness of our results if the singularities are as well hinted to be avoided in this alternative approach.

According to our previous work \cite{Bouhmadi-Lopez:2014jfa}, the Friedmann equations of the metric $g_{\mu\nu}$ and the auxiliary metric $q_{\mu\nu}$ near the big rip singularity when $a\rightarrow\infty$ are:
\begin{equation}
\begin{split}
\kappa H^2=\kappa\Big(\frac{\dot{a}}{a}\Big)^2\approx\frac{4\sqrt{|w|^3}}{3(3w+1)^2}a^{\epsilon} \,,
\qquad
\kappa H_q^2=\kappa\Big(\frac{1}{b}\frac{db}{d{\tilde t}}\Big)^2\approx\frac{1}{3} \,.
\end{split}
\end{equation}
The new variable $\tilde{t}$ is defined as $\tilde{t}\equiv\sqrt{U}t$ and corresponds to the rescaled time such that the auxiliary metric can be written in a FLRW form \cite{Bouhmadi-Lopez:2014jfa}. By reminding that in GR, the Friedmann equation indeed corresponds to the Hamiltonian, we can then use these Friedmann equations as effective Hamiltonian $\mathcal{H}_g$ and $\mathcal{H}_q$, with their canonical variables being $(a,p_a)$ and $(b,p_b)$, respectively, where the conjugate momenta $p_a$ and $p_b$ are defined by (this definition is simply inspired in GR)
\begin{equation}
\begin{split}
p_a\equiv-6a\dot{a} \,,
\qquad
p_b\equiv-6b\frac{db}{d\tilde t} \,.
\end{split}
\end{equation}
Note that we have set $8\pi G=1$. Afterwards, we assume that the total wave function of the Universe $\Psi(a,b)$ satisfies the WDW equation constructed as the product of $\mathcal{H}_g$ and $\mathcal{H}_q$:
\begin{equation}
\mathcal{H}_t\Psi(a,b)=\mathcal{H}_g\mathcal{H}_q\Psi(a,b)=0.
\end{equation}
We obtain a ``phenomenological'' WDW equation:
\begin{equation}
\Big(\frac{\partial^2}{\partial a^2}+\frac{48\sqrt{|w^3|}}{\kappa\hbar^2(3w+1)^2}a^{4+\epsilon}\Big)\Big(\frac{\partial^2}{\partial b^2}+\frac{12}{\kappa\hbar^2}b^4\Big)\Psi(a,b)=0.
\end{equation}
We next use the ansatz $\Psi(a,b)=\psi_1(a)\psi_2(b)$ and obtain
\begin{equation}
\Big(\frac{d^2}{d a^2}+\frac{48\sqrt{|w^3|}}{\kappa\hbar^2(3w+1)^2}a^{4+\epsilon}\Big)\psi_1(a)\Big(\frac{d^2}{db^2}+\frac{12}{\kappa\hbar^2}b^4\Big)\psi_2(b)=0,
\label{0}
\end{equation}
which means that at least one of the following equations must hold
\begin{subequations}\label{11221122}
\begin{align}
\Big(\frac{d^2}{d a^2}+\frac{48\sqrt{|w^3|}}{\kappa\hbar^2(3w+1)^2}a^{4+\epsilon}\Big)\psi_1(a)&=0,\label{1}\\
\Big(\frac{d^2}{db^2}+\frac{12}{\kappa\hbar^2}b^4\Big)\psi_2(b)&=0.\label{2}
\end{align}
\end{subequations}

\begin{itemize}
\item If Eq.~\eqref{1} holds, the solution is \cite{mathhandbook}
\begin{equation}
\psi_1(a)=C_1\sqrt{a}J_{\nu}(A_2 a^{\frac{1}{2\nu}})+C_2\sqrt{a}Y_{\nu}(A_2 a^{\frac{1}{2\nu}}),
\label{bessel1}
\end{equation}
where
\begin{equation}
\begin{split}
A_2\equiv2\nu\Big[\frac{48\sqrt{|w^3|}}{\kappa\hbar^2(3w+1)^2}\Big]^{\frac{1}{2}}\,,
\qquad
\nu\equiv\frac{1}{6+\epsilon}\,.
\end{split}
\end{equation}
In the appendix~\ref{app}, we will show that the solution $\psi_1(a)$ in Eq.~\eqref{bessel1} approaches zero when $a\rightarrow\infty$.

Aside from Eq.~\eqref{1}, the left hand side of Eq.~\eqref{2} should be bounded to ensure the validity of the solution \eqref{0}. Therefore, we have:
\begin{equation}
\Big(\frac{d^2}{db^2}+\frac{12}{\kappa\hbar^2}b^4\Big)\psi_2(b)=K_1(b),
\label{K1d}
\end{equation}
where $K_1(b)$ is an arbitrary bounded function of $b$. In the appendix~\ref{app}, we will also prove that $\psi_2(b)$ is bounded for all values of $b$. Therefore, we can conclude that the total wave function $\psi_1(a)\psi_2(b)$ vanishes when $a\rightarrow\infty$ for all $b$, and the big rip singularity is expected to be avoided.

\item On the other hand, if Eq.~\eqref{2} holds, the solution of $\psi_2(b)$ is \cite{mathhandbook}
\begin{equation}
\psi_2(b)=C_1\sqrt{b}J_{1/6}(A_3b^3)+C_2\sqrt{b}Y_{1/6}(A_3b^3),
\label{bessel2}
\end{equation}
where
\begin{equation}
A_3\equiv\Big(\frac{4}{\kappa\hbar^2}\Big)^{\frac{1}{2}}.
\end{equation}
In the appendix~\ref{app}, we will show that the solution $\psi_2(b)$ in Eq.~\eqref{bessel2} is bounded for all values of $b$. Furthermore, the left hand side of Eq.~\eqref{1} should be bounded to ensure the validity of Eq.~\eqref{0}. We then have
\begin{equation}
\Big(\frac{d^2}{d a^2}+\frac{48\sqrt{|w^3|}}{\kappa\hbar^2(3w+1)^2}a^{4+\epsilon}\Big)\psi_1(a)=K_2(a),
\label{K2d}
\end{equation}
where $K_2(a)$ is an arbitrary bounded function of $a$. In the appendix~\ref{app}, we will also prove that the solution $\psi_1(a)$ of Eq.~\eqref{K2d} will vanish when $a$ approaches infinity. Then we can conclude that the total wave function $\psi_1(a)\psi_2(b)$ vanishes when $a\rightarrow\infty$ for all $b$, and the big rip singularity is expected to be avoided.
\end{itemize}

\section{Conclusion}\label{sec:conclusion}
Although the EiBI theory is characterized by its ability to cure the big bang singularity at the early universe \cite{Banados:2010ix,Scargill:2012kg}, it has been shown that the big rip singularity is unavoidable in the EiBI phantom model \cite{Bouhmadi-Lopez:2013lha,Bouhmadi-Lopez:2014jfa}. It is then natural to ask whether some quantum effects can prevent the occurrence of this fatal cosmic doomsday. Despite the lack of a full understanding of quantum gravity, quantum cosmology in which a homogeneous, isotropic and spatially flat universe is assumed can comparably reduce the complexity to tackle this problem \cite{qgkiefer}. In this work, we consider a quantum geometrodynamical approach, in which the WDW equation plays a central role, to investigate whether the big rip singularity in the EiBI model can be avoided. The WDW equation determines the evolution of the wave function of the universe as a whole. As usual, we use DW condition, which states that the wave function vanishes at the region corresponding to the classical singularities, as a hint to judge the avoidance of the cosmological singularity of interest. It is worth to be mentioned that even though similar issues have been widely studied in the framework of GR, it is the first time that this approach is applied to a modified theory of gravity constructed upon a Palatini formalism. The independence of the connection and metric, in addition to the complicated square root structure in the EiBI action make the construction of the WDW equation troublesome. 

To overcome these problems, we consider an alternative action proposed in Ref.~\cite{Delsate:2012ky}, which is dynamically equivalent to the original EiBI action, to construct the classical Hamiltonian. This action is linear in the Ricci scalar purely defined by the auxiliary metric, and does not contain square roots of the spacetime curvatures. It can be regarded as an equivalent interpretation based on the Einstein frame of the original EiBI framework. After obtaining a suitable Hamiltonian, we construct the WDW equation under two different factor ordering choices. Exact solutions can be derived under a specific assumption that the auxiliary scale factor $b$ is related to the physical scale factor $a$ through their classical equations of motion. For general circumstances in which these quantities are treated independently, we use a WKB approximation and argue that the wave function vanishes on the region corresponding to the classical big rip singularity for all values of $b$. According to the DW condition, the big rip singularity is expected to be cured in this approach. The two factor ordering choices lead to the same conclusion. 

On the other hand, we also construct an effective WDW equation by simply promoting the classical Friedmann equations near the big rip singularity. The definitions of the conjugate momenta are inspired by GR. In this \textit{handwaving} approach, we find that the total wave function vanishes near the configuration corresponding to the classical big rip for all values of the auxiliary scale factor $b$. Therefore, the singularity is expected to be avoided according to the DW condition. The conclusions of the different approaches considered in this work are consistent.

At this point we would like to emphasize that the DW condition that we regard as a guiding hint for the singularity avoidance within the quantum realm is not strictly sufficient to avoid the classical singularity. The validity of this criterium is tightly dependent on the existence of square integrable functions, and therefore on a consistent probability interpretation for the wave function. In any case, it is the probability amplitudes for wave packets that should vanish close to the region of configuration space corresponding to the classical singularity or abrupt event. The problem is the fact that these square integrable functions require an appropriate Hilbert space associated with its inner product which defines a proper measure factor. It is not obvious that this can always be done in a straightforward way in quantum cosmology. In addition, our argumentation has to be taken with great care as in Refs.~\cite{Barvinsky:1993jf,Kamenshchik:2012ij,Barvinsky:2013aya}, it was shown that in some particular cases while the wave function vanishes at the classical singularity, the probability associated with that event does not vanish as a consequence of the fact that the Faddeev-Popov measure used to define the probability blows up at this point. However, as we mentioned previously, we can still regard the DW condition as a potential hint for singularity avoidance. Furthermore, this work can be seen as a pioneering work to generalize the associated quantum geometrodynamical treatments to a modified theory of gravity constructed within the Palatini formalism.

As we have just proven, the existence of an alternative action, which takes the form of a linear structure like that in GR, can reduce the complexity of constructing the WDW equation in this modified theory of gravity. For other theories formulated within the Palatini formalism or even within the metric affine theory, the existence of such an alternative action can shed some light on further investigating the quantum cosmology of these theories and GR itself. Furthermore, it is also interesting to see whether other cosmological singularities unavoidable in EiBI theory can be avoided in a similar way. We will tackle these interesting issues on the near future.

\acknowledgments

The authors acknowledge Pisin Chen for comments and discussions on this work. The work of MBL is supported by the Portuguese Agency ``Funda\c{c}\~{a}o para a Ci\^{e}ncia e Tecnologia'' through an Investigador FCT Research contract, with reference IF/01442/2013/CP1196/CT0001. She also wishes to acknowledge the partial support from the Basque government Grant No.~IT592-13 (Spain) and FONDOS FEDER under grant FIS2014-57956-P (Spanish government). This research work is supported by the Portuguese grand UID/MAT/00212/2013. C.-Y.C. is supported by Taiwan National Science Council under Project No. NSC 97-2112-M-002-026-MY3 and by Taiwans National Center for Theoretical Sciences (NCTS).

\appendix
\section{Derivation of the EiBI field equations}\label{app0}
In this appendix, we will obtain the EiBI field equations by varying the action \eqref{actioneibi}
\begin{equation}
S_{EiBI}=\frac{2}{\kappa}\int d^4x\Big[\sqrt{|g_{\mu\nu}+\kappa R_{\mu\nu}(\Gamma)|}-\lambda\sqrt{-g}\Big]+S_M(g),
\end{equation}
based on the Palatini formalism, i.e., the variation will be carried with respect to $\Gamma$ and $g$. By varying the action with respect to the independent metric and the connection, we obtain
\begin{eqnarray}
\delta{S}_{EiBI}&=&\frac{1}{\kappa}\int d^4x\Big[\sqrt{|\hat{g}+\kappa \hat{R}|}[(\hat{g}+\kappa \hat{R})^{-1}]^{\mu\nu}-\lambda\sqrt{-g}g^{\mu\nu}\Big]\delta g_{\mu\nu}+\delta{S}_M(g)\nonumber\\
&&+\int d^4x\sqrt{|\hat{g}+\kappa \hat{R}|}[(\hat{g}+\kappa \hat{R})^{-1}]^{\mu\nu}\delta R_{\mu\nu}(\Gamma),
\label{app1.2}
\end{eqnarray}
where a hat denotes the tensor nature of a given object without making explicit reference to the tensor components. Imposing $\delta S_{EiBI}=0$ in Eq.~\eqref{app1.2} and reminding that the metric and the connection are independent variables, the field equation of the metric $g$ reads
\begin{equation}
\sqrt{|\hat{g}+\kappa \hat{R}|}[(\hat{g}+\kappa \hat{R})^{-1}]^{\mu\nu}=\lambda\sqrt{-g}g^{\mu\nu}-\sqrt{-g}\kappa T^{\mu\nu}=0,
\end{equation}
i.e., Eq.~\eqref{eq2}. Note that the energy momentum tensor $T^{\mu\nu}$ is defined in Eq.~\eqref{energymomentumT}.
On the other hand, the variation with respect to the connection can be written as
\begin{eqnarray}
&&\int d^4x\sqrt{|\hat{g}+\kappa \hat{R}|}[(\hat{g}+\kappa \hat{R})^{-1}]^{\mu\nu}\delta R_{\mu\nu}(\Gamma)\nonumber\\
&=&\int d^4x\sqrt{|\hat{g}+\kappa \hat{R}|}[(\hat{g}+\kappa \hat{R})^{-1}]^{\mu\nu}(\nabla_{\lambda}\delta\Gamma^\lambda_{\mu\nu}-\nabla_{\nu}\delta\Gamma^{\lambda}_{\lambda\mu})\nonumber\\
&=&-\int d^4x\delta\Gamma^\sigma_{\alpha\mu}\Big\{\delta^\alpha_\nu\nabla_{\sigma}\Big[\sqrt{|\hat{g}+\kappa \hat{R}|}[(\hat{g}+\kappa \hat{R})^{-1}]^{\mu\nu}\Big]-\delta^\alpha_\sigma\nabla_{\nu}\Big[\sqrt{|\hat{g}+\kappa \hat{R}|}[(\hat{g}+\kappa \hat{R})^{-1}]^{\mu\nu}\Big]\Big\}.\nonumber
\end{eqnarray}
To obtain the last line an integration by part has been used. Note that $\nabla$ is the covariant derivative constructed from the connection. After imposing that the above variation vanishes and after taking a trace on $\alpha$ and $\sigma$, the field equation of the connection reads
\begin{equation}
\nabla_{\alpha}\Big\{\sqrt{|\hat{g}+\kappa \hat{R}|}[(\hat{g}+\kappa \hat{R})^{-1}]^{\mu\nu}\Big\}=0.
\label{app1.4}
\end{equation}
Then, we introduce the covariant derivative of the tensor density $\sqrt{|g_{\mu\nu}+\kappa R_{\mu\nu}|}$ and the covariant derivative of $[(\hat{g}+\kappa \hat{R})^{-1}]^{\mu\nu}$ as follows
\begin{eqnarray}
\nabla_{\alpha}\sqrt{|\hat{g}+\kappa \hat{R}|}&=&\partial_{\alpha}\sqrt{|\hat{g}+\kappa \hat{R}|}-\sqrt{|\hat{g}+\kappa \hat{R}|}\Gamma^\rho_{\rho\alpha}\nonumber\\
&=&-\sqrt{|\hat{g}+\kappa \hat{R}|}\Big\{\frac{1}{2}(\hat{g}+\kappa \hat{R})_{\rho\sigma}\partial_\alpha[(\hat{g}+\kappa \hat{R})^{-1}]^{\rho\sigma}+\Gamma^\rho_{\rho\alpha}\Big\},\nonumber\\
\nabla_{\alpha}[(\hat{g}+\kappa \hat{R})^{-1}]^{\mu\nu}&=&\partial_\alpha[(\hat{g}+\kappa \hat{R})^{-1}]^{\mu\nu}+\Gamma^{\mu}_{\alpha\beta}[(\hat{g}+\kappa \hat{R})^{-1}]^{\nu\beta}+\Gamma^{\nu}_{\alpha\beta}[(\hat{g}+\kappa \hat{R})^{-1}]^{\mu\beta},\nonumber
\end{eqnarray}
where the equation,
\begin{equation}
\partial_{\alpha}\sqrt{|\hat{g}+\kappa \hat{R}|}=-\frac{1}{2}\sqrt{|\hat{g}+\kappa \hat{R}|}(\hat{g}+\kappa \hat{R})_{\rho\sigma}\partial_\alpha[(\hat{g}+\kappa \hat{R})^{-1}]^{\rho\sigma},
\end{equation} 
is used to derive the second line of the first equation.

After contracting Eq.~\eqref{app1.4} with $(\hat{g}+\kappa \hat{R})_{\mu\nu}$ it can be shown that
\begin{equation}
\nabla_{\alpha}\sqrt{|\hat{g}+\kappa \hat{R}|}=0.
\label{app1.6}
\end{equation} 
Finally, from Eqs.~\eqref{app1.4} and \eqref{app1.6} we obtain
\begin{equation}
\nabla_{\alpha}[(\hat{g}+\kappa \hat{R})^{-1}]^{\mu\nu}=0.
\end{equation}
Therefore, we can define the auxiliary metric $q_{\mu\nu}$, which is compatible with the connection, such as the auxiliary metric that satisfies $\lambda q_{\mu\nu}=g_{\mu\nu}+\kappa R_{\mu\nu}$, i.e., Eq.~\eqref{eq1}.

\section{The phase of the wave function within a WKB approximation}\label{app1}
\subsection{Rewriting the phase of the wave function}
The integral \eqref{integrals}, that defines the phase of the wave function \eqref{solution}, can be written as
\begin{equation}
I=\frac{a^3}{\sqrt{2}}\int\frac{1}{\xi}\sqrt{\xi^3-\frac{3}{2}\xi^2+\frac{K^2(a)}{2}}d\xi\equiv\frac{a^3}{\sqrt{2}}\int\frac{1}{\xi}\sqrt{P(\xi)}d\xi
\label{ourintegralori}
\end{equation}
after making a change of variable $a^2\xi=\textit{e}^{2x}$. Note that $K(a)=\lambda+\kappa\rho(a)$ is larger than one because we only focus on a universe with a positive effective cosmological constant $\Lambda=(\lambda-1)/\kappa$. The discriminant $D$ of the cubic polynomial $P(\xi)$ reads \cite{mathhandbook}
\begin{equation}
D\equiv Q^3+\mathcal{R}^2=\frac{1}{16}[K^4(a)-K^2(a)],
\end{equation}
where $Q=-1/4$ and $\mathcal{R}=(1-2K^2(a))/8$. It can be seen that $D$ is positive because $K(a)$ is larger than one as mentioned a few lines above. Therefore $P(\xi)$ has only one real root. The roots are \cite{mathhandbook}
\begin{equation}
\begin{split}
\xi_1=\frac{1}{2}-\cosh{\frac{\theta}{3}}\,,
\qquad
\xi_2=a_r+ib_r\,,
\qquad
\xi_3=\xi_2^*=a_r-ib_r\,,
\end{split}
\end{equation}
where $\theta$, $a_r$ and $b_r$ are defined as
\begin{equation}
\begin{split}
\cosh{\theta}=\frac{-\mathcal{R}}{\sqrt{-Q^3}}\,,
\qquad
\sinh{\theta}=\sqrt{\frac{D}{-Q^3}}\,,
\qquad
a_r\equiv\frac{1}{2}\Big(1+\cosh{\frac{\theta}{3}}\Big)\,,
\qquad
b_r\equiv\frac{\sqrt{3}}{2}\sinh{\frac{\theta}{3}}\,.
\end{split}
\end{equation}
Because $\cosh{(\theta/3)\ge1}$, the real root $\xi_1$ is negative and $a_r$ is positive.

Finally, the integral \eqref{ourintegralori} can be written as
\begin{equation}
I=\frac{a^3}{\sqrt{2}}\int\frac{1}{\xi}\sqrt{(\xi-\xi_1)[(\xi-a_r)^2+b_r^2]}d\xi.
\label{xitozeta}
\end{equation}
Under the redefinition $\zeta=\xi-a_r$ and $c\equiv a_r-\xi_1>0$, \eqref{xitozeta} can be further expressed as
\begin{equation}
I=\frac{a^3}{\sqrt{2}}\int\frac{R_1(\zeta)}{\sqrt{(\zeta+c)(\zeta^2+b_r^2)}}d\zeta,
\label{zetaintegral}
\end{equation}
where $R_1(\zeta)$ is a rational function of $\zeta$: 
\begin{equation}
R_1(\zeta)=\frac{(\zeta+c)(\zeta^2+b_r^2)}{\zeta+a_r}.
\label{R1y}
\end{equation}

\subsection{Canonical form of the elliptic integral}
To calculate the integral \eqref{zetaintegral} analytically, we need to rewrite it in its canonical form which can be expressed solely by three kinds of elliptic integrals \cite{elliptic}. To reach that goal we proceed as follows \cite{elliptic}:
First, we define $S_1\equiv \zeta^2+b_r^2$ and $S_2\equiv \zeta+c$. Then we find a constant $\lambda_i$ which makes $S_1-\lambda_i S_2$ a perfect square. This requires that the discriminant of the quadratic polynomial $S_1-\lambda_i S_2$ to be zero. The constant $\lambda_i$ has two solutions
\begin{equation}
\begin{split}
\lambda_1=2\sqrt{b_r^2+c^2}-2c\,,
\qquad
\lambda_2=-2\sqrt{b_r^2+c^2}-2c\,,
\end{split}
\end{equation}
and each of them satisfies
\begin{equation}
\begin{split}
S_1-\lambda_1 S_2=(\zeta-\alpha)^2\,,
\qquad
S_1-\lambda_2 S_2=(\zeta+\beta)^2\,,
\end{split}
\end{equation}
where $\alpha=\lambda_1/2$ and $\beta=-\lambda_2/2$. Therefore, $S_1$ and $S_2$ can be rewritten as
\begin{equation}
\begin{split}
S_1=\frac{(\zeta-\alpha)^2-(\zeta+\beta)^2}{\lambda_2-\lambda_1}\,,
\qquad
S_2=\frac{\lambda_2(\zeta-\alpha)^2-\lambda_1(\zeta+\beta)^2}{\lambda_2-\lambda_1}\,.
\end{split}
\end{equation}
Substituting the previous equations into the integral \eqref{zetaintegral} and making a further change of variable: $u=(\zeta-\alpha)/(\zeta+\beta)$, the integral \eqref{zetaintegral} can be written as
\begin{equation}
I=\frac{a^3}{\sqrt{2}}\int\frac{R_1(u)du}{(\alpha+\beta)\sqrt{A\Big(1-u^2\Big)\Big(1+\frac{|\lambda_2|}{|\lambda_1|}u^2\Big)}}\,,
\end{equation}
where
\begin{equation*}
A=\frac{\lambda_1}{(\lambda_2-\lambda_1)^2}.
\end{equation*}
Note that the original integration interval of $\xi$ is within $(0,\infty)$. The changes of variables $\zeta=\xi-a_r$ and $u=(\zeta-\alpha)/(\zeta+\beta)$ imply that the integration interval of $u$ is within $(u_i,1)$ where $-1<u_i<0$ and $u$ is, in addition, an increasing function of $\zeta$ on that interval. The latter can be seen by evaluating $du/d\zeta$ and reminding that $\alpha$ and $\beta$ are positive. For the sake of simplicity, we will consider the integration interval of $u$ such that $u\in(0,u)$. An arbitrary initial value of the integration only relates to the initial phase of the oscillation thus it does not change our conclusion significantly.

For any rational function $R_1(u)$, we can always define $R_2(u^2)$ and $R_3(u^2)$ such that
\begin{equation}
\begin{split}
2R_2(u^2)\equiv R_1(u)+R_1(-u)\,,
\qquad
2uR_3(u^2)\equiv R_1(u)-R_1(-u)\,,
\end{split}
\end{equation}
then
\begin{equation}
R_1(u)=R_2(u^2)+uR_3(u^2).
\label{decompose}
\end{equation}
Therefore, the integral can be written as
\begin{equation}
I=\frac{a^3}{\sqrt{2A\frac{|\lambda_2|}{|\lambda_1|}}(\alpha+\beta)}\Bigg\{\int\frac{R_2(u^2)du}{\sqrt{\Big(1-u^2\Big)\Big(\frac{|\lambda_1|}{|\lambda_2|}+u^2\Big)}}+\int\frac{uR_3(u^2)du}{\sqrt{\Big(1-u^2\Big)\Big(\frac{|\lambda_1|}{|\lambda_2|}+u^2\Big)}}\Bigg\}.
\end{equation}
The second integral can be integrated using elementary functions. On the other hand, the first integral can be decomposed by partial fractions and can be written as a combination of three kinds of elliptic integrals \cite{elliptic}. 

By changing the variable from $\zeta$ to $u$ and according to Eq.~\eqref{R1y}, we have
\begin{equation}
R_1(u)=\frac{2(b_r^2+c^2)(1+u)(\alpha+\beta u^2)}{[\alpha+a_r+(\beta-a_r)u](1-u)^2}.
\end{equation}
Following Eq.~\eqref{decompose}, we can obtain
\begin{equation}
R_2(u^2)=2(b_r^2+c^2)\Big[a_0+\frac{4}{(1-u^2)^2}+\frac{a_1}{(1-u^2)}+\frac{a_2}{(p^2-u^2)}\Big],
\end{equation}
where 
\begin{equation}
\begin{split}
p&=\frac{\alpha+a_r}{\beta-a_r}\,,
\qquad
a_0=\frac{\beta}{\beta-a_r}\,,
\qquad
a_1=-\frac{2a_r+3\alpha+5\beta}{\alpha+\beta}\,,
\\
a_2&=-\frac{(\alpha-\beta+2a_r)(\alpha\beta+a_r^2)(\alpha+a_r)}{(\beta-a_r)^3(\alpha+\beta)}\,.
\end{split}
\end{equation}
Similarly, the anti-symmetric part of $R_1(u)$ reads
\begin{equation}
uR_3(u^2)=2(b_r^2+c^2)u\Big[\frac{4}{(1-u^2)^2}+\frac{b_1}{(1-u^2)}+\frac{b_2}{(p^2-u^2)}\Big],
\end{equation}
where
\begin{equation}
\begin{split}
b_1=-\frac{2a_r+\alpha+3\beta}{\alpha+\beta}\,,
\qquad
b_2=\frac{(\alpha-\beta+2a_r)(\alpha\beta+a_r^2)}{(\beta-a_r)^2(\alpha+\beta)}\,.
\end{split}
\end{equation}
According to the following integrals \cite{integral}
\begin{eqnarray}
I_1&=&b_1\int_0^u\frac{udu}{\sqrt{(1-u^2)^3(L^2+u^2)}}=\frac{b_1}{1+L^2}\sqrt{\frac{L^2+u^2}{1-u^2}}\Bigg|^u_0,\nonumber\\
I_2&=&\int_0^u\frac{4udu}{\sqrt{(1-u^2)^5(L^2+u^2)}}=\frac{4(3+L^2-2u^2)}{3(1+L^2)^2}\sqrt{\frac{L^2+u^2}{(1-u^2)^3}}\Bigg|^u_0,\nonumber\\
I_3&=&b_2\int_0^u\frac{udu}{(p^2-u^2)\sqrt{(1-u^2)(L^2+u^2)}}\nonumber\\&=&\frac{b_2}{\sqrt{(1-p^2)(L^2+p^2)}}\tanh^{-1}{\sqrt{\frac{(L^2+p^2)(1-u^2)}{(1-p^2)(L^2+u^2)}}}\Bigg|^u_0,\nonumber
\end{eqnarray}
and
\begin{eqnarray}
I_4&=&a_0\int_0^u\frac{du}{\sqrt{(1-u^2)(L^2+u^2)}}=\frac{a_0}{\sqrt{1+L^2}}F(\gamma,r),\nonumber\\
I_5&=&a_1\int_0^u\frac{du}{\sqrt{(1-u^2)^3(L^2+u^2)}}=\frac{a_1}{\sqrt{1+L^2}}[F(\gamma,r)-E(\gamma,r)]+\frac{a_1u}{\sqrt{(L^2+u^2)(1-u^2)}},\nonumber\\
I_6&=&\int_0^u\frac{4du}{\sqrt{(1-u^2)^5(L^2+u^2)}}\nonumber\\
&=&\frac{4[(2L^2+3)F(\gamma,r)-(2L^2+4)E(\gamma,r)]}{3\sqrt{(L^2+1)^3}}+\frac{4u[3L^2+4-(2L^2+3)u^2]}{3(L^2+1)\sqrt{(L^2+u^2)(1-u^2)^3}},\nonumber\\
I_7&=&a_2\int_0^u\frac{du}{(p^2-u^2)\sqrt{(1-u^2)(L^2+u^2)}}\nonumber\\
&=&\frac{a_2}{p^2(p^2+L^2)\sqrt{L^2+1}}\Big[L^2\Pi\Big(\gamma,\frac{p^2+L^2}{p^2(L^2+1)},r\Big)+p^2F(\gamma,r)\Big],\nonumber
\end{eqnarray}
the result of the integration \eqref{ourintegralori} is 
\begin{equation}
I=C(a)\mathcal{I}(u)\equiv C(a)(I_1+I_2+I_3+I_4+I_5+I_6+I_7),\label{resutsint}
\end{equation}
where 
\begin{equation*}
\begin{split}
L^2=\frac{|\lambda_1|}{|\lambda_2|}\,,
\qquad
r=\frac{1}{\sqrt{L^2+1}}\,,
\qquad
\gamma=\sin^{-1}{\Big(u\sqrt{\frac{L^2+1}{L^2+u^2}}\Big)}\,,
\end{split}
\end{equation*}
and
\begin{equation}
C(a)=\frac{4(b_r^2+c^2)a^3}{\sqrt{2|\lambda_2|}},\label{Caaa}
\end{equation}
where on the above equations $F(\gamma,r)$, $E(\gamma,r)$ and $\Pi [\gamma,(p^2+L^2)/p^2(L^2+1),r]$ are the elliptic integral of first, second and third kind, respectively \cite{integral}.

\begin{figure}[t]
\centering
\graphicspath{{fig/}}
\includegraphics[scale=1.2]{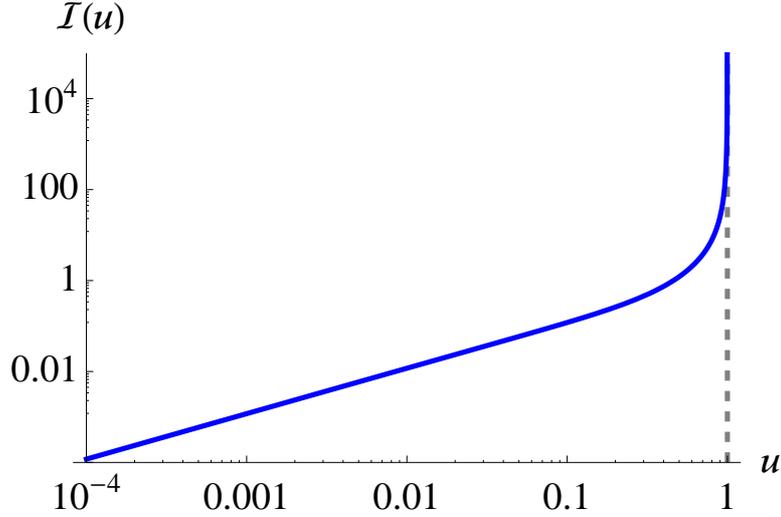}
\caption{The result of the integration $\mathcal{I}(u)$ where $I\equiv C(a)\mathcal{I}(u)$ when the scale factor $a$ is very large. The definitions of $\mathcal{I}(u)$ and $C(a)$ are given in Eqs.~\eqref{resutsint} and \eqref{Caaa}, respectively. Note that $C(a)\propto(\lambda+\kappa\rho(a))a^3$ when $a$ is large.}
\label{plot}
\end{figure}

Near the classical big rip singularity where the scale factor $a$ as well as $K(a)$ become large, we have
\begin{equation}
\begin{split}
\cosh{\theta}\approx2K^2(a)\,,
\qquad
\cosh{\frac{\theta}{3}}\approx\Big(\frac{K^2(a)}{2}\Big)^{\frac{1}{3}}\,,
\end{split}
\end{equation}
and
\begin{equation}
C(a)\approx\frac{6}{\sqrt{2\sqrt{3}+3}}K(a)a^3.
\end{equation}
In this limit, the behavior of $\mathcal{I}(u)$ is shown in figure.~\ref{plot}. Furthermore, because $C(a)\propto(\lambda+\kappa\rho(a))a^3$, the result of the integral $I$ goes to positive infinity when $a$ is very large. This corresponds to a rapidly oscillating wave function under a WKB approximation as described in subsection \ref{wkbsec}.

\section{Boundness and asymptotic behavior of the wave functions in the phenomenological approach}\label{app}
We start considering the inhomogeneous differential equation:
\begin{equation}
\Big(\frac{d^2}{dz^2}+\frac{C^2}{4m^2}z^{\frac{1}{m}-2}\Big)\psi(z)=\mathcal{K}(z),
\label{diff}
\end{equation}
where $C$ is a positive constant and $0<m\le 1/6$. The source term $\mathcal{K}(z)$ is a bounded function of $z$. Therefore, one can always choose a positive constant $\mathcal{K}$ such that $|\mathcal{K}(z)|\le\mathcal{K}<\infty$ for all $z$. Note that Eqs.~\eqref{K1d} and \eqref{K2d} correspond to $m=1/6$ and $0<m<1/6$, respectively. The solution to this inhomogeneous differential equation can be written as a combination of the homogeneous solution and a particular solution:
\begin{equation}
\psi(z)=\psi_h(z)+\psi_p(z),
\end{equation}
where the homogeneous part $\psi_h(z)$ is \cite{mathhandbook}
\begin{eqnarray}
\psi_h(z)&=&C_1\sqrt{z}J_{m}(Cz^{\frac{1}{2m}})+C_2\sqrt{z}Y_{m}(Cz^{\frac{1}{2m}})\nonumber\\
&=&D_1\sqrt{z}J_{m}(Cz^{\frac{1}{2m}})+D_2\sqrt{z}J_{-m}(Cz^{\frac{1}{2m}}).
\label{homo}
\end{eqnarray}
We next remind the following inequality (see Eq.~9.1.62 of Ref.~\cite{mathhandbook})
\begin{equation}
|J_\mu(z)|\le\frac{|\frac{1}{2}z|^{\mu}}{\Gamma(\mu+1)},
\label{ineq}
\end{equation}
where $\mu\ge-1/2$ and $z$ is real, and apply it to the solution \eqref{homo}, so that we obtain
\begin{equation}
|\psi_h(z)|\le\frac{|D_1||\frac{C}{2}|^m}{\Gamma(1+m)}z+\frac{|D_2||\frac{C}{2}|^{-m}}{\Gamma(1-m)}.
\end{equation}
Therefore, $\psi_h(z)$ is bounded for $0\le z<\infty$. Note that $0<m\le 1/6$.

Furthermore, when $z\rightarrow\infty$ the homogeneous solution can be approximated as \cite{mathhandbook}
\begin{equation}
\psi_h(z)\approx\sqrt{\frac{2}{\pi C}}z^{\frac{1}{2}-\frac{1}{4m}}\Big[C_1\cos{\Big(Cz^{\frac{1}{2m}}-\frac{1}{2}m\pi-\frac{1}{4}\pi\Big)}+C_2\sin{\Big(Cz^{\frac{1}{2m}}-\frac{1}{2}m\pi-\frac{1}{4}\pi\Big)}\Big].
\end{equation}
It can be proven that $\psi_h(z)\rightarrow 0$ when $z\rightarrow\infty$ because $0<m\le 1/6$. Therefore, we can conclude that the homogeneous solution $\psi_h(z)$, which is the solution to the differential equation \eqref{diff} with $\mathcal{K}(z)=0$, approaches zero when $z\rightarrow\infty$ and it is bounded for all values of $z$.

The particular solution to the inhomogeneous differential equation \eqref{diff} can be obtained by using the method of variation of parameters \cite{differentialbook}:
\begin{equation}
\psi_p(z)=\sqrt{z}J_{m}(Cz^{\frac{1}{2m}})u_1(z)+\sqrt{z}J_{-m}(Cz^{\frac{1}{2m}})u_2(z),
\end{equation}
where $u_1(z)$ and $u_2(z)$ satisfy
\begin{eqnarray}
\sqrt{z}\Big[\frac{du_1}{dz}J_{m}(Cz^{\frac{1}{2m}})+\frac{du_2}{dz}J_{-m}(Cz^{\frac{1}{2m}})\Big]&=&0,\nonumber\\
\frac{du_1}{dz}\frac{d}{dz}\Big[\sqrt{z}J_{m}(Cz^{\frac{1}{2m}})\Big]+\frac{du_2}{dz}\frac{d}{dz}\Big[\sqrt{z}J_{-m}(Cz^{\frac{1}{2m}})\Big]&=&\mathcal{K}(z).\nonumber
\end{eqnarray}
After some calculations, we obtain \cite{mathhandbook}
\begin{equation}
\begin{split}
\frac{du_1}{dz}=\Big[\frac{m\pi}{\sin{(m\pi)}}\Big]\mathcal{K}(z)\psi_{h-}(z)\,,
\qquad
\frac{du_2}{dz}=-\Big[\frac{m\pi}{\sin{(m\pi)}}\Big]\mathcal{K}(z)\psi_{h+}(z)\,,
\end{split}
\end{equation}
where 
\begin{equation*}
\begin{split}
\psi_{h-}(z)\equiv\sqrt{z}J_{-m}(Cz^{\frac{1}{2m}})\,,
\qquad
\psi_{h+}(z)\equiv\sqrt{z}J_{m}(Cz^{\frac{1}{2m}})\,.
\end{split}
\end{equation*}
Therefore, the particular solution $\psi_p(z)$ can be written as
\begin{equation}
\psi_{p}(z)=\Big[\frac{m\pi}{\sin{(m\pi)}}\Big]\bigg\{\psi_{h+}(z)\int^z\mathcal{K}(z')\psi_{h-}(z')dz'-\psi_{h-}(z)\int^z\mathcal{K}(z')\psi_{h+}(z')dz'\bigg\}.
\label{p1}
\end{equation}
Note that the integration constants in \eqref{p1} can be absorbed into the homogeneous solutions, so they can be neglected in the integration. By considering the absolute value $|\psi_p(z)|$ and reminding that $|\mathcal{K}(z)|\le\mathcal{K}<\infty$ for all $z$, we have
\begin{eqnarray}
&&|\psi_p(z)|\nonumber\\
&=&\Big|\frac{m\pi}{\sin{(m\pi)}}\Big|\bigg|\psi_{h+}(z)\int^z\mathcal{K}(z')\psi_{h-}(z')dz'-\psi_{h-}(z)\int^z\mathcal{K}(z')\psi_{h+}(z')dz'\bigg|\nonumber\\
&\le&\Big|\frac{m\pi}{\sin{(m\pi)}}\Big|\Bigg\{\bigg|\psi_{h+}(z)\int^z\mathcal{K}(z')\psi_{h-}(z')dz'\bigg|+\bigg|\psi_{h-}(z)\int^z\mathcal{K}(z')\psi_{h+}(z')dz'\bigg|\Bigg\}\nonumber\\
&\le&\Big|\frac{m\mathcal{K}\pi}{\sin{(m\pi)}}\Big|\Bigg\{\bigg|\psi_{h+}(z)\bigg|\int^z\bigg|\psi_{h-}(z')\bigg|dz'+\bigg|\psi_{h-}(z)\bigg|\int^z\bigg|\psi_{h+}(z')\bigg|dz'\Bigg\}\nonumber\\
&\le&\Big|\frac{m\mathcal{K}\pi}{\sin{(m\pi)}}\Big|\frac{3z^2}{2\Gamma(1+m)\Gamma(1-m)}.\nonumber
\end{eqnarray}
Note that the inequality \eqref{ineq} has been used to obtain the last inequality. In summary, it can be seen that the particular solution $\psi_p(z)$ is bounded for $0\le z<\infty$.

When $z\rightarrow\infty$, we can always choose an integration interval in Eq.~\eqref{p1} in which the particular solution can be approximated as follows
\begin{equation*}
\psi_{p}(z)\Big|_{z\rightarrow\infty}\approx\tilde{C}\Bigg\{\tilde{\psi}_{h+}(z)\int^z\mathcal{K}(z')\tilde{\psi}_{h-}(z')dz'-\tilde{\psi}_{h-}(z)\int^z\mathcal{K}(z')\tilde{\psi}_{h+}(z')dz'\Bigg\},
\label{p2}
\end{equation*}
where $\tilde{C}\equiv m\pi/\sin{(m\pi)}$, and 
\begin{equation}
\tilde{\psi}_{h\pm}(z)\equiv\Big[\psi_{h\pm}(z)\Big]_{z\rightarrow\infty}\approx\sqrt{\frac{2}{\pi C}}z^{\frac{1}{2}-\frac{1}{4m}}\cos{\Big(Cz^{\frac{1}{2m}}\mp\frac{m}{2}\pi-\frac{\pi}{4}\Big)}.
\end{equation}
Finally, we have
\begin{eqnarray}
\Big|\psi_{p}(z)\Big|_{z\rightarrow\infty}
&\approx&\tilde{C}\bigg|\tilde{\psi}_{h+}(z)\int^z\mathcal{K}(z')\tilde{\psi}_{h-}(z')dz'-\tilde{\psi}_{h-}(z)\int^z\mathcal{K}(z')\tilde{\psi}_{h+}(z')dz'\bigg|\nonumber\\
&\le&\tilde{C}\Bigg\{\Big|\tilde{\psi}_{h+}(z)\Big|\bigg|\int^z\mathcal{K}(z')\tilde{\psi}_{h-}(z')dz'\bigg|+\Big|\tilde{\psi}_{h-}(z)\Big|\bigg|\int^z\mathcal{K}(z')\tilde{\psi}_{h+}(z')dz'\bigg|\Bigg\}\nonumber\\
&\le&\tilde{C}\mathcal{K}\Bigg\{\Big|\tilde{\psi}_{h+}(z)\Big|\int^z\Big|\tilde{\psi}_{h-}(z')\Big|dz'+\Big|\tilde{\psi}_{h-}(z)\Big|\int^z\Big|\tilde{\psi}_{h+}(z')\Big|dz'\Bigg\}\nonumber\\
&\le&\tilde{C}\mathcal{K}\sqrt{\frac{2}{\pi C}}\bigg[\Big|\tilde{\psi}_{h+}(z)\Big|+\Big|\tilde{\psi}_{h-}(z)\Big|\bigg]\int^zz'^{\frac{1}{2}-\frac{1}{4m}}dz'\nonumber\\
&\le&\frac{8\mathcal{K}m}{C\sin{(m\pi)}}\Big(\frac{1}{2m}-3\Big)^{-1}z^{2-\frac{1}{2m}}\rightarrow 0,
\end{eqnarray}
when $0<m<1/6$. Note that $2-1/(2m)<0$ in this case.

If $m=1/6$, we have
\begin{eqnarray}
\Big|\psi_{p}(z)\Big|_{z\rightarrow\infty}
&\le&\tilde{C}\mathcal{K}\sqrt{\frac{2z}{\pi C}}\bigg[\Big|J_{1/6}(Cz^3)\Big|+\Big|J_{-1/6}(Cz^3)\Big|\bigg]_{z\rightarrow\infty}\times\int^z\frac{dz'}{z'}\nonumber\\
&\le&\frac{4\mathcal{K}}{3C}\frac{\ln{z}}{z}\rightarrow 0.
\end{eqnarray}
Therefore, the particular solution $\psi_{p}(z)$ as well as the total solution $\psi(z)$ approach zero when $z\rightarrow\infty$. Furthermore, it can also be safely said that the total solution $\psi(z)$ is bounded for all values of $z$. 

Consequently, in this appendix we have proven that:
\begin{itemize}
\item The solutions to the differential equations \eqref{1} and \eqref{K2d} vanish when $a\rightarrow\infty$.

\item The solutions to the differential equations \eqref{2} and \eqref{K1d} are bounded for all values of $b$.
\end{itemize}


\begin{thebibliography}{99}

\bibitem{Delsate:2012ky} 
  T.~Delsate and J.~Steinhoff, \emph{New insights on the matter-gravity coupling paradigm}, 
  \emph{Phys.\ Rev.\ Lett.\ } {\bf 109} (2012) 021101
  [\href{http://arxiv.org/abs/1201.4989v3}{arXiv:1201.4989v3}].

\bibitem{gravitation} 
  C.~W.~Misner, K.~S.~Thorne, and J.~A.~Wheeler, \emph{Gravitation}, (W.~H.~Freeman, 1973).

\bibitem{largescale} 
  S.~W.~Hawking, G.~F.~R.~Ellis, \emph{The Large Scale Structure of Space-Time}, (Cambridge University Press, 1973).

\bibitem{Perlmutter:1998np} 
  S.~Perlmutter {\it et al.} [Supernova Cosmology Project Collaboration],
  \emph{Measurements of Omega and Lambda from 42 high redshift supernovae},
  \emph{Astrophys.\ J.\ } {\bf 517} (1999) 565 [\href{http://arxiv.org/abs/astro-ph/9812133}{astro-ph/9812133}].

\bibitem{Riess:1998cb} 
  A.~G.~Riess {\it et al.} [Supernova Search Team Collaboration],
  \emph{Observational evidence from supernovae for an accelerating universe and a cosmological constant},
  \emph{Astron.\ J.\ } {\bf 116} (1998) 1009 [\href{http://arxiv.org/abs/astro-ph/9805201}{astro-ph/9805201}].

\bibitem{Nojiri:2005sx} 
  S.~Nojiri, S.~D.~Odintsov and S.~Tsujikawa,
  \emph{Properties of singularities in (phantom) dark energy universe},
  \emph{Phys.\ Rev.\ D} {\bf 71} (2005) 063004 [\href{http://arxiv.org/abs/hep-th/0501025}{hep-th/0501025}].

\bibitem{Starobinsky:1999yw} 
  A.~A.~Starobinsky,
  \emph{Future and origin of our universe: Modern view},
  \emph{Grav.\ Cosmol.\ } {\bf 6} (2000) 157 [\href{http://arxiv.org/abs/astro-ph/9912054}{astro-ph/9912054}].

\bibitem{Caldwell:1999ew} 
  R.~R.~Caldwell,
  \emph{A Phantom menace?},
  \emph{Phys.\ Lett.\ B} {\bf 545} (2002) 23 [\href{http://arxiv.org/abs/astro-ph/9908168}{astro-ph/9908168}].

\bibitem{Caldwell:2003vq}
  R.~R.~Caldwell, M.~Kamionkowski and N.~N.~Weinberg,
  \emph{Phantom energy and cosmic doomsday},
  \emph{Phys.\ Rev.\ Lett.\ } {\bf 91} (2003) 071301 [\href{http://arxiv.org/abs/astro-ph/0302506}{astro-ph/0302506}].

\bibitem{Carroll:2003st} 
  S.~M.~Carroll, M.~Hoffman and M.~Trodden,
  \emph{Can the dark energy equation - of - state parameter w be less than -1?},
  \emph{Phys.\ Rev.\ D} {\bf 68} (2003) 023509 [\href{http://arxiv.org/abs/astro-ph/0301273}{astro-ph/0301273}].
  
\bibitem{Chimento:2003qy} 
  L.~P.~Chimento and R.~Lazkoz,
  \emph{On the link between phantom and standard cosmologies},
  \emph{Phys.\ Rev.\ Lett.\ } {\bf 91} (2003) 211301 [\href{http://arxiv.org/abs/gr-qc/0307111}{gr-qc/0307111}].
  
\bibitem{Dabrowski:2003jm} 
  M.~P.~D\c{a}browski, T.~Stachowiak and M.~Szyd{\l }owski,
  \emph{Phantom cosmologies},
  \emph{Phys.\ Rev.\ D} {\bf 68} (2003) 103519 [\href{http://arxiv.org/abs/hep-th/0307128}{hep-th/0307128}].

\bibitem{GonzalezDiaz:2003rf} 
  P.~F.~Gonz\'{a}lez-D\'{i}az,
  \emph{K-essential phantom energy: Doomsday around the corner?},
  \emph{Phys.\ Lett.\ B} {\bf 586} (2004) 1 [\href{http://arxiv.org/abs/astro-ph/0312579}{astro-ph/0312579}].

\bibitem{GonzalezDiaz:2004vq} 
  P.~F.~Gonz\'{a}lez-D\'{i}az,
  \emph{Axion phantom energy},
  \emph{Phys.\ Rev.\ D} {\bf 69} (2004) 063522 [\href{http://arxiv.org/abs/hep-th/0401082}{hep-th/0401082}].

\bibitem{BouhmadiLopez:2004me}
  M.~Bouhmadi-L\'{o}pez and J.~A.~Jim\'{e}nez Madrid,
  \emph{Escaping the big rip?},
  \emph{JCAP} {\bf 0505} (2005) 005 [\href{http://arxiv.org/abs/astro-ph/0404540}{astro-ph/0404540}].

\bibitem{BouhmadiLopez:2006fu} 
  M.~Bouhmadi-L\'{o}pez, P.~F.~Gonz\'{a}lez-D\'{i}az and P.~Mart\'{i}n-Moruno,
  \emph{Worse than a big rip?},
  \emph{Phys.\ Lett.\ B} {\bf 659} (2008) 1 [\href{http://arxiv.org/abs/gr-qc/0612135}{gr-qc/0612135}].

\bibitem{Banados:2010ix} 
  M.~Ba\~{n}ados and P.~G.~Ferreira,
  \emph{Eddington's theory of gravity and its progeny},
  \emph{Phys.\ Rev.\ Lett.\ } {\bf 105} (2010) 011101 [\href{http://arxiv.org/abs/1006.1769v2}{arXiv:1006.1769v2}]
  Erratum: [\emph{Phys.\ Rev.\ Lett.\ } {\bf 113} (2014) no. 11, 119901].

\bibitem{Deser:1998rj}
  S.~Deser and G.~W.~Gibbons,
  \emph{Born-Infeld-Einstein actions?},
  \emph{Class.\ Quant.\ Grav.\ } {\bf 15} (1998) L35
   [\href{http://arxiv.org/abs/hep-th/9803049}{hep-th/9803049}].

\bibitem{Pani:2011mg} 
  P.~Pani, V.~Cardoso and T.~Delsate,
  \emph{Compact stars in Eddington inspired gravity},
  \emph{Phys.\ Rev.\ Lett.\ } {\bf 107} (2011) 031101 [\href{http://arxiv.org/abs/1106.3569}{arXiv:1106.3569}].

\bibitem{Casanellas:2011kf} 
  J.~Casanellas, P.~Pani, I.~Lopes and V.~Cardoso,
  \emph{Testing alternative theories of gravity using the Sun},
  \emph{Astrophys.\ J.\ } {\bf 745} (2012) 15 [\href{http://arxiv.org/abs/1109.0249}{arXiv:1109.0249}].

\bibitem{Avelino:2012ge} 
  P.~P.~Avelino,
  \emph{Eddington-inspired Born-Infeld gravity: astrophysical and cosmological constraints},
  \emph{Phys.\ Rev.\ D} {\bf 85} (2012) 104053 [\href{http://arxiv.org/abs/1201.2544}{arXiv:1201.2544}].

\bibitem{Pani:2012qb} 
  P.~Pani, T.~Delsate and V.~Cardoso,
  \emph{Eddington-inspired Born-Infeld gravity. Phenomenology of non-linear gravity-matter coupling},
  \emph{Phys.\ Rev.\ D} {\bf 85} (2012) 084020 [\href{http://arxiv.org/abs/1201.2814}{arXiv:1201.2814}].

\bibitem{EscamillaRivera:2012vz} 
  C.~Escamilla-Rivera, M.~Ba\~{n}ados and P.~G.~Ferreira,
  \emph{A tensor instability in the Eddington inspired Born-Infeld Theory of Gravity},
  \emph{Phys.\ Rev.\ D} {\bf 85} (2012) 087302 [\href{http://arxiv.org/abs/1204.1691}{arXiv:1204.1691}].

\bibitem{Avelino:2012ue} 
  P.~P.~Avelino and R.~Z.~Ferreira,
  \emph{Bouncing Eddington-inspired Born-Infeld cosmologies: an alternative to Inflation ?},
  \emph{Phys.\ Rev.\ D} {\bf 86} (2012) 041501 [\href{http://arxiv.org/abs/1205.6676}{arXiv:1205.6676}].

\bibitem{Avelino:2012qe} 
  P.~P.~Avelino,
  \emph{Eddington-inspired Born-Infeld gravity: nuclear physics constraints and the validity of the continuous fluid approximation},
  \emph{JCAP} {\bf 1211} (2012) 022 [\href{http://arxiv.org/abs/1207.4730}{arXiv:1207.4730}].

\bibitem{Cho:2012vg} 
  I.~Cho, H.~C.~Kim and T.~Moon,
  \emph{Universe Driven by Perfect Fluid in Eddington-inspired Born-Infeld Gravity},
  \emph{Phys.\ Rev.\ D} {\bf 86} (2012) 084018 [\href{http://arxiv.org/abs/1208.2146v1}{arXiv:1208.2146v1}][arXiv:1208.2146v1].

\bibitem{Pani:2012qd} 
 P.~Pani and T.~P.~Sotiriou,
  \emph{Surface singularities in Eddington-inspired Born-Infeld gravity},
\emph{Phys.\ Rev.\ Lett.\ } {\bf 109} (2012) 251102 [\href{http://arxiv.org/abs/1209.2972v2}{arXiv:1209.2972v2}].

\bibitem{Scargill:2012kg} 
  J.~H.~C.~Scargill, M.~Ba\~{n}ados and P.~G.~Ferreira,
  \emph{Cosmology with Eddington-inspired Gravity},
  \emph{Phys.\ Rev.\ D} {\bf 86} (2012) 103533 [\href{http://arxiv.org/abs/1210.1521}{arXiv:1210.1521}].

\bibitem{Cho:2013usa} 
  I.~Cho and H.~C.~Kim,
  \emph{New synthesis of matter and gravity: A nongravitating scalar field},
  \emph{Phys.\ Rev.\ D} {\bf 88} (2013) 064038 [\href{http://arxiv.org/abs/1302.3341v2}{arXiv:1302.3341v2}].

\bibitem{Bouhmadi-Lopez:2013lha} 
  M.~Bouhmadi-L\'{o}pez, C.~-Y.~Chen and P.~Chen,
  \emph{Is Eddington-Born-Infeld theory really free of cosmological singularities?},
  \emph{Eur.\ Phys.\ J.\ C} {\bf 74} (2014) 2802 [\href{http://arxiv.org/abs/1302.5013}{arXiv:1302.5013}].

\bibitem{Cho:2013pea} 
  I.~Cho, H.~C.~Kim and T.~Moon,
  \emph{Precursor of Inflation},
  \emph{Phys.\ Rev.\ Lett.\ } {\bf 111} (2013) 071301 [\href{http://arxiv.org/abs/1305.2020v1}{arXiv:1305.2020v1}].

\bibitem{Harko:2013wka} 
  T.~Harko, F.~S.~N.~Lobo, M.~K.~Mak and S.~V.~Sushkov,
  \emph{Structure of neutron, quark and exotic stars in Eddington-inspired Born-Infeld gravity},
  \emph{Phys.\  Rev.\ D} {\bf 88} (2013) 044032 [\href{http://arxiv.org/abs/1305.6770v1}{arXiv:1305.6770v1}].

\bibitem{Harko:2013aya} 
  T.~Harko, F.~S.~N.~Lobo, M.~K.~Mak and S.~V.~Sushkov,
  \emph{Wormhole geometries in Eddington-Inspired Born-Infeld gravity},
  \emph{Mod.\ Phys.\ Lett.\ A} {\bf 30} (2015) no. 35, 1550190 [\href{http://arxiv.org/abs/1307.1883v1}{arXiv:1307.1883v1}].

\bibitem{Yang:2013hsa} 
  K.~Yang, X.~-L.~Du and Y.~-X.~Liu,
  \emph{Linear perturbations in Eddington-inspired Born-Infeld gravity},
  \emph{Phys.\ Rev.\ D} {\bf 88} (2013) 124037 [\href{http://arxiv.org/abs/1307.2969}{arXiv:1307.2969}].

\bibitem{Olmo:2013gqa}
  G.~J.~Olmo, D.~Rubiera-Garcia and H.~Sanchis-Alepuz,
  \emph{Geonic black holes and remnants in Eddington-inspired Born-Infeld gravity},
  \emph{Eur.\ Phys.\ J.\ C} {\bf 74} (2014) 2804 [\href{http://arxiv.org/abs/1311.0815}{arXiv:1311.0815}].

\bibitem{Sham:2013cya} 
  Y.~-H.~Sham, L.~-M.~Lin and P.~T.~Leung,
  \emph{Testing universal relations of neutron stars with a nonlinear matter-gravity coupling theory},
  \emph{Astrophys.\ J.\ } {\bf 781} (2014) 66 [\href{http://arxiv.org/abs/1312.1011}{arXiv:1312.1011}].

\bibitem{Du:2014jka} 
  X.~L.~Du, K.~Yang, X.~H.~Meng and Y.~X.~Liu,
  \emph{Large Scale Structure Formation in Eddington-inspired Born-Infeld Gravity},
  \emph{Phys.\ Rev.\ D} {\bf 90} (2014) 044054 [\href{http://arxiv.org/abs/1403.0083v2}{arXiv:1403.0083v2}].

\bibitem{Makarenko:2014lxa} 
  A.~N.~Makarenko, S.~Odintsov and G.~J.~Olmo,
  \emph{Born-Infeld-$f(R)$ gravity},
  \emph{Phys.\ Rev.\ D} {\bf 90} (2014) 024066 [\href{http://arxiv.org/abs/1403.7409v2}{arXiv:1403.7409v2}].

\bibitem{Makarenko:2014nca} 
  A.~N.~Makarenko, S.~D.~Odintsov and G.~J.~Olmo,
  \emph{Little Rip, $\Lambda$CDM and singular dark energy cosmology from Born-Infeld-$f(R)$ gravity},
\emph{Phys.\ Lett.\  B} {\bf 734} (2014) 36 [\href{http://arxiv.org/abs/1404.2850}{arXiv:1404.2850}].

\bibitem{Wei:2014dka} 
  S.~W.~Wei, K.~Yang and Y.~X.~Liu,
  \emph{Black hole solution and strong gravitational lensing in Eddington-inspired Born-Infeld gravity},
  \emph{Eur.\ Phys.\ J.\ C} {\bf 75} (2015) 253 [\href{http://arxiv.org/abs/1405.2178v3}{arXiv:1405.2178v3}]
  Erratum: [\emph{Eur.\ Phys.\ J.\ C} {\bf 75} (2015) 331].

\bibitem{Odintsov:2014yaa} 
  S.~D.~Odintsov, G.~J.~Olmo and D.~Rubiera-Garcia,
  \emph{Born-Infeld gravity and its functional extensions},
  \emph{Phys.\ Rev.\ D} {\bf 90} (2014) 044003 [\href{http://arxiv.org/abs/1406.1205v2}{arXiv:1406.1205v2}].

\bibitem{Bouhmadi-Lopez:2014jfa} 
  M.~Bouhmadi-L\'{o}pez, C.~Y.~Chen and P.~Chen,
  \emph{Eddington-Born-Infeld cosmology: a cosmographic approach, a tale of doomsdays and the fate of bound structures},
  \emph{Eur.\ Phys.\ J.\ C} {\bf 75} (2015) no. 2, 90 [\href{http://arxiv.org/abs/1406.6157}{arXiv:1406.6157}].

\bibitem{Bouhmadi-Lopez:2014tna} 
  M.~Bouhmadi-L\'{o}pez, C.~Y.~Chen and P.~Chen,
  \emph{Cosmological singularities in Born-Infeld determinantal gravity},
  \emph{Phys.\ Rev.\ D} {\bf 90} (2014) no. 12, 123518 [\href{http://arxiv.org/abs/1407.5114}{arXiv:1407.5114}].

\bibitem{Jimenez:2014fla} 
  J.~B.~Jim\'{e}nez, L.~Heisenberg and G.~J.~Olmo,
  \emph{Infrared lessons for ultraviolet gravity: the case of massive gravity and Born-Infeld},
  \emph{JCAP} {\bf 1411} (2014) 004 [\href{http://arxiv.org/abs/1409.0233v2}{arXiv:1409.0233v2}].

\bibitem{Makarenko:2014cca} 
  A.~N.~Makarenko, S.~D.~Odintsov, G.~J.~Olmo and D.~Rubiera-Garcia,
  \emph{Early-time cosmic dynamics in $f(R)$ and $f(|\hat\Omega|)$ extensions of Born-Infeld gravity},
  \emph{TSPU Bulletin} {\bf 12} (2014) 158 [\href{http://arxiv.org/abs/1411.6193}{arXiv:1411.6193}].

\bibitem{Sotani:2014lua} 
  H.~Sotani and U.~Miyamoto,
  \emph{Properties of an electrically charged black hole in Eddington-inspired Born-Infeld gravity},
  \emph{Phys.\ Rev.\ D} {\bf 90} (2014) no. 12, 124087 [\href{http://arxiv.org/abs/1412.4173}{arXiv:1412.4173}].

\bibitem{Cho:2014xaa} 
  I.~Cho and N.~K.~Singh,
  \emph{Scalar perturbation produced at the pre-inflationary stage in Eddington-inspired Born-Infeld gravity},
  \emph{Eur.\ Phys.\ J.\ C} {\bf 75} (2015) no. 6, 240 [\href{http://arxiv.org/abs/1412.6344v2}{arXiv:1412.6344v2}].

\bibitem{Potapov:2014iva} 
  A.~A.~Potapov, R.~Izmailov, O.~Mikolaychuk, N.~Mikolaychuk, M.~Ghosh and K.~K.~Nandi,
  \emph{Constraint on dark matter central density in the Eddington inspired Born-Infeld (EiBI) gravity with input from Weyl gravity},
  \emph{JCAP} {\bf 1507} (2015) no. 07, 018 [\href{http://arxiv.org/abs/1412.7897v6}{arXiv:1412.7897v6}].

\bibitem{Sotani:2015tya} 
  H.~Sotani,
  \emph{Magnetized relativistic stellar models in Eddington-inspired Born-Infeld gravity},
  \emph{Phys.\ Rev.\ D} {\bf 91} (2015) no. 8, 084020 [\href{http://arxiv.org/abs/1503.07942}{arXiv:1503.07942}].

\bibitem{Bambi:2015sla}
  C.~Bambi, G.~J.~Olmo and D.~Rubiera-Garcia,
  \emph{Melvin Universe in Born-Infeld gravity},
  \emph{Phys.\ Rev.\ D} {\bf 91} (2015) no.10, 104010 [\href{http://arxiv.org/abs/1504.01827}{arXiv:1504.01827}].

\bibitem{Jana:2015cha} 
  S.~Jana and S.~Kar,
  \emph{Born-Infeld gravity coupled to Born-Infeld electrodynamics},
  \emph{Phys.\ Rev.\ D} {\bf 92} (2015) 084004 [\href{http://arxiv.org/abs/1504.05842v3}{arXiv:1504.05842v3}].

\bibitem{Olmo:2015dba}
  G.~J.~Olmo, D.~Rubiera-Garcia and A.~Sanchez-Puente,
  \emph{Classical resolution of black hole singularities via wormholes},
  \emph{Eur.\ Phys.\ J.\ C} {\bf 76} (2016) no.3, 143 [\href{http://arxiv.org/abs/1504.07015}{arXiv:1504.07015}].

\bibitem{Shaikh:2015oha} 
  R.~Shaikh,
  \emph{Lorentzian wormholes in Eddington-inspired Born-Infeld gravity},
  \emph{Phys.\ Rev.\ D} {\bf 92} (2015) 024015 [\href{http://arxiv.org/abs/1505.01314v2}{arXiv:1505.01314v2}].
        
\bibitem{Izmailov:2015xsa} 
  R.~Izmailov, A.~A.~Potapov, A.~I.~Filippov, M.~Ghosh and K.~K.~Nandi,
  \emph{Upper limit on the central density of dark matter in the Eddington-inspired Born-Infeld (EiBI) gravity},
  \emph{Mod.\ Phys.\ Lett.\ A} {\bf 30} (2015) no. 11, 1550056 [\href{http://arxiv.org/abs/1506.04023v1}{arXiv:1506.04023v1}].

\bibitem{Chen:2015eha} 
  C.~Y.~Chen, M.~Bouhmadi-L\'{o}pez and P.~Chen,
  \emph{Modified Eddington-inspired-Born-Infeld Gravity with a Trace Term},
  \emph{Eur.\ Phys.\ J.\ C} {\bf 76} (2016) 40 [\href{http://arxiv.org/abs/1507.00028v3}{arXiv:1507.00028v3}].

\bibitem{Sotani:2015ewa} 
  H.~Sotani and U.~Miyamoto,
  \emph{Strong gravitational lensing by an electrically charged black hole in Eddington-inspired Born-Infeld gravity},
  \emph{Phys.\ Rev.\ D} {\bf 92} (2015) no. 4, 044052 [\href{http://arxiv.org/abs/1508.03119v1}{arXiv:1508.03119v1}].

\bibitem{Olmo:2015bya}
  G.~J.~Olmo, D.~Rubiera-Garcia and A.~Sanchez-Puente,
  \emph{Geodesic completeness in a wormhole spacetime with horizons},
  \emph{Phys.\ Rev.\ D} {\bf 92} (2015) no.4, 044047 [\href{http://arxiv.org/abs/1508.03272}{arXiv:1508.03272}].

\bibitem{Tamang:2015tmd} 
  A.~Tamang, A.~A.~Potapov, R.~Lukmanova, R.~Izmailov and K.~K.~Nandi,
  \emph{On the generalized wormhole in the Eddington-inspired Born-Infeld gravity},
  \emph{Class.\ Quant.\ Grav.\ } {\bf 32} (2015) no. 23, 235028 [\href{http://arxiv.org/abs/1512.01451v1}{arXiv:1512.01451v1}].

\bibitem{Elizalde:2016vsd} 
  E.~Elizalde and A.~N.~Makarenko,
  \emph{Singular inflation from Born-Infeld-f(R) gravity}, [\href{http://arxiv.org/abs/1606.05211}{arXiv:1606.05211}].
  
\bibitem{Bambi:2016xme}
  C.~Bambi, D.~Rubiera-Garcia and Y.~Wang,
  \emph{Black hole solutions in functional extensions of Born-Infeld gravity}, [\href{http://arxiv.org/abs/1608.04873}{arXiv:1608.04873}].

\bibitem{qgkiefer}
C.~Kiefer, \emph{Quantum Gravity}. Second edition (Oxford University Press, Oxford, 2007).

\bibitem{DeWitt:1967yk} 
  B.~S.~DeWitt,
  \emph{Quantum Theory of Gravity. 1. The Canonical Theory},
  \emph{Phys.\ Rev.\ } {\bf 160} (1967) 1113.

\bibitem{Dabrowski:2006dd} 
  M.~P.~D\c{a}browski, C.~Kiefer and B.~Sandh\"{o}fer,
  \emph{Quantum phantom cosmology},
  \emph{Phys.\ Rev.\ D} {\bf 74} (2006) 044022 [\href{http://arxiv.org/abs/hep-th/0605229v1}{hep-th/0605229v1}].

\bibitem{Kamenshchik:2007zj} 
  A.~Kamenshchik, C.~Kiefer and B.~Sandh\"{o}fer,
  \emph{Quantum cosmology with big-brake singularity},
  \emph{Phys.\ Rev.\ D} {\bf 76} (2007) 064032 [\href{http://arxiv.org/abs/0705.1688v2}{arXiv:0705.1688v2}].

\bibitem{BouhmadiLopez:2009pu} 
  M.~Bouhmadi-L\'{o}pez, C.~Kiefer, B.~Sandh\"{o}fer and P.~Vargas Moniz,
  \emph{On the quantum fate of singularities in a dark-energy dominated universe},
  \emph{Phys.\ Rev.\ D} {\bf 79} (2009) 124035 [\href{http://arxiv.org/abs/0905.2421v2}{arXiv:0905.2421v2}].

\bibitem{Kamenshchik:2013naa}
  A.~Y.~Kamenshchik,
  \emph{Quantum cosmology and late-time singularities},
  \emph{Class.\ Quant.\ Grav.\ } {\bf 30} (2013) 173001 [\href{http://arxiv.org/abs/1307.5623}{arXiv:1307.5623}].

\bibitem{Bouhmadi-Lopez:2013tua} 
  M.~Bouhmadi-L\'{o}pez, C.~Kiefer and M.~Kr\"{a}mer,
  \emph{Resolution of type IV singularities in quantum cosmology},
  \emph{Phys.\ Rev.\ D} {\bf 89} (2014) no. 6, 064016 [\href{http://arxiv.org/abs/1312.5976v2}{arXiv:1312.5976v2}].

\bibitem{Albarran:2015tga}
  I.~Albarran and M.~Bouhmadi-L\'opez,
  \emph{Quantisation of the holographic Ricci dark energy model},
  \emph{JCAP} {\bf 1508} (2015) no.08,  051 [\href{http://arxiv.org/abs/1505.01353v2}{arXiv:1505.01353v2}].

\bibitem{Albarran:2015cda}
  I.~Albarran, M.~Bouhmadi-L\'opez, F.~Cabral and P.~Mart\'{i}n-Moruno,
  \emph{The quantum realm of the "Little Sibling" of the Big Rip singularity},
  \emph{JCAP} {\bf 1511} (2015) no.11,  044 [\href{http://arxiv.org/abs/1509.07398v1}{arXiv:1509.07398v1}].

\bibitem{Albarran:2016ewi} 
  I.~Albarran, M.~Bouhmadi-L\'{o}pez, C.~Kiefer, J.~Marto and P.~Vargas Moniz,
  \emph{Classical and quantum cosmology of the little rip abrupt event}, [\href{http://arxiv.org/abs/1604.08365}{arXiv:1604.08365}].

\bibitem{Shojai:2008er} 
  A.~Shojai and F.~Shojai,
  \emph{f(R) Quantum Cosmology},
  \emph{Gen.\ Rel.\ Grav.\ } {\bf 40} (2008) 1967 [\href{http://arxiv.org/abs/0801.3496v1}{arXiv:0801.3496v1}].

\bibitem{Vakili:2008uj} 
  B.~Vakili,
  \emph{Noether symmetric f(R) quantum cosmology and its classical correlations},
  \emph{Phys.\ Lett.\ B} {\bf 669} (2008) 206 [\href{http://arxiv.org/abs/0809.4591v1}{arXiv:0809.4591v1}].

\bibitem{Vakili:2009he} 
  B.~Vakili,
  \emph{Quadratic quantum cosmology with Schutz' perfect fluid},
  \emph{Class.\ Quant.\ Grav.\ } {\bf 27} (2010) 025008 [\href{http://arxiv.org/abs/0908.0998v2}{arXiv:0908.0998v2}].

\bibitem{Kamenshchik:2016rtr} 
  A.~Kamenshchik, C.~Kiefer and N.~Kwidzinski,
  \emph{Classical and quantum cosmology of Born-Infeld type models},
  \emph{Phys.\ Rev.\ D} {\bf 93} (2016) no. 8, 083519 [\href{http://arxiv.org/abs/1602.01319v2}{arXiv:1602.01319v2}].





  


\bibitem{capozziellogravity}
S.~Capozziello and V. Faraoni, \emph{Beyond Einstein Gravity}, Springer, New York, U.S.A. (2011).

\bibitem{mathhandbook}
M.~Abramowitz and I.~Stegun, \emph{Handbook on Mathematical Functions} (Dover, 1980).

\bibitem{mathhandbook2}
J.~Mathews and R.~L.~Walker, \emph{Mathematical Methods of Physics} (California Institute of Technology, 1969).

\bibitem{Barvinsky:1993jf} 
  A.~O.~Barvinsky,
  \emph{Unitarity approach to quantum cosmology},
  \emph{Phys.\ Rept.\ } {\bf 230} (1993), 237.

\bibitem{Kamenshchik:2012ij} 
  A.~Y.~Kamenshchik and S.~Manti,
  \emph{Classical and quantum Big Brake cosmology for scalar field and tachyonic models},
  \emph{Phys.\ Rev.\ D} {\bf 85} (2012) 123518 [\href{http://arxiv.org/abs/1202.0174}{arXiv:1202.0174}].

\bibitem{Barvinsky:2013aya}
  A.~O.~Barvinsky and A.~Y.~Kamenshchik,
  \emph{Selection rules for the Wheeler-DeWitt equation in quantum cosmology},
  \emph{Phys.\ Rev.\ D} {\bf 89} (2014) no.4, 043526 [\href{http://arxiv.org/abs/1312.3147}{arXiv:1312.3147}].

    
\bibitem{elliptic}
E.~T.~Whittaker and G.~N.~Watson, \emph{A Course of Modern Analysis}, Cambridge University Press; 4th edition (1927)

\bibitem{integral}
I.~S.~Gradshteyn and I.~M.~Ryzhik, \emph{Tables of Integrals, Series, and Products}, Academic Press Inc (1965)

\bibitem{differentialbook}
William F.~Trench, \emph{Elementary Differential Equations} (2013). Faculty Authored Books. Book 8  

\end{thebibliography}
\end{document}